\documentclass[a4paper,11pt]{article}
\usepackage{jheppub}  
\usepackage{amsmath}
\usepackage{amssymb} 
\usepackage{graphicx}
\usepackage{orcidlink}
\usepackage{multirow}

\usepackage{physics}
\usepackage{mathrsfs}

\usepackage[caption=false,labelformat=simple]{subfig}

\newcommand{\lumi}{\ensuremath{\mathscr{L}}}
\newcommand{\operatorO}{\ensuremath{\hat{\mathcal{O}}}}
\newcommand{\expectationO}{\ensuremath{\expval{\mathcal{O}}}}
\newcommand{\bellOperator}{\ensuremath{\hat{\mathcal{B}}}}

\newcommand{\Ithree}{\ensuremath{\mathcal{I}_3}}

\newcommand{\jth}{\ensuremath{j^{\rm th}}}

\preprint{IMSc/2025/07}

\title{Unfolding quantum entanglement from $h\to ZZ^*\to jj\ell\ell$ at a muon collider}

\author[a]{Songshaptak De\,\orcidlink{0000-0003-3174-7425}}
\author[b]{Atri Dey\,\orcidlink{0000-0002-1645-7641}}
\author[c]{Tousik Samui\,\orcidlink{0000-0002-1485-6155}}
\affiliation[a]{Institute of Physics, Bhubaneswar, Sachivalaya Marg,
Sainik School, Bhubaneswar 751005, India}
\affiliation[b]{Department of Physics and Astronomy, Uppsala University,
Box 516, SE-751 20 Uppsala, Sweden}
\affiliation[c]{The Institute of Mathematical Sciences, IV Cross Road, CIT Campus, Taramani, Chennai 600113, India}

\emailAdd{deswaptak@gmail.com}
\emailAdd{atri.dey@physics.uu.se}
\emailAdd{atridey1993@gmail.com}
\emailAdd{tousiks@imsc.res.in}
\emailAdd{tousiksamui@gmail.com}

\abstract{We explore the potential to study quantum entanglement through Bell-type inequalities in Higgs boson decays at a future muon collider. Our analysis focuses on the channel $\mu^+ \mu^- \to \nu \bar{\nu} h \to \nu \bar{\nu} ZZ^*$, with one $Z$ decaying to charged leptons and the other decaying hadronically into jets.
We study the violation of the CGLMP inequality using the optimal Bell operator for the bipartite qutrit system from $h \to ZZ^*$. The entanglement measure $\Ithree$ is constructed from spin-correlated angular observables of the $Z$ decay products. An unfolding method on the angular variables is applied to correct for hadronization and detector effects, recovering the advantage of the hadronic mode with higher event yield and reduced uncertainty.
The study is performed at 1, 3, and 10 TeV centre-of-mass energies, assuming 10 ab$^{-1}$ integrated luminosity for each case. At 1 TeV, we use a boosted decision tree for signal isolation, while at higher energies, simple cut-based analyses are sufficient. We find clear Bell inequality violation with the expected values $\Ithree = 2.625\pm0.012$, $2.623\pm0.004$, and $2.582\pm0.010$ for the 1, 3, and 10 TeV machines, respectively. Overall, a strong level of entanglement close to the maximum achievable value of 2.9149 for a two-qutrit system 
can be measured with very small uncertainties due to the large event yield in the hadronic mode.}


\begin{document}
\maketitle

\section{Introduction}\label{sec:intro}
Quantum mechanics (QM) is equipped with a very special feature
called entanglement, which states that a quantum mechanical system
containing two or more subsystems cannot be described as a simple
product of two independent subsystems. In other words, they are
inseparable in nature and carry the footprint of the interactions the
system has undergone in the past. The subsystems remain connected
through an overall wave function even when they are spatially separated.
This phenomenon was first proposed by Einstein, Podolsky, and
Rosen\,\cite{Einstein:1935rr}. Such connections lead to correlations
between the subsystems that cannot be explained by any means of
classical mechanics\,\cite{Nielsen:2012yss}. These correlations are often observed through
spin-spin correlations among the subsystems.

While entanglement is easily explained and observed in non-relativistic
systems described by QM\,\cite{Nielsen:2012yss}, the same property is not as straightforward in
systems involving particles with high momentum\,\cite{Peres:2002ip,Czachor:1996cj}. Such systems require a
relativistic treatment consistent with Lorentz transformations, as well
as the principles of quantum theory. The framework describing these
systems is quantum field theory (QFT). In relativistic motion, the
degree of spin entanglement can be reduced or enhanced depending on the
reference frame connected by Lorentz transformations\,\cite{Peres:2002ip,Gingrich:2002ota,Terashima:2002pth,Bergou:2003nni}. The spin wave
function may exhibit different degrees of quantum correlation when
transformed from one inertial frame to another related by a Lorentz
boost\,\cite{Gingrich:2002ota,Terashima:2002pth}. Therefore, although QFT is built upon Lorentz invariance, it does
not automatically guarantee that quantum entanglement, arising from its
quantum nature, remains invariant. However, it has been shown that while
the entanglement between spin and momentum changes under Lorentz boosts,
the total entanglement of the entire wave function remains
invariant\,\cite{Gingrich:2002ota,Bergou:2003nni}.

Hence, the treatment of entanglement within QFT requires a different
formalism to properly define and study quantum correlations in such
systems. This makes it highly relevant to test entanglement in collider
experiments\,\cite{Fabbrichesi:2023cev,Aoude:2023hxv,Barr:2024djo}, where interactions are governed by QFT. At the same time,
testing such phenomena in colliders is non-trivial. Several formalisms
have been developed to study entanglement in different systems, and it
is therefore important to investigate its validity across different
energy scales, since varying energies involve different Lorentz boosts.
Furthermore, possible complications arising from effects such as
renormalization and QCD non-perturbative dynamics should also be
examined to understand whether entanglement can be meaningfully observed
in such systems.

In recent years, there has been a growing effort to apply concepts from
quantum information theory to collider physics. This effort is
especially relevant in the context of offering new ways to characterize
correlations between particles produced in high-energy
collisions\,\cite{Aoude:2023hxv,Barr:2024djo,Ding:2025mzj}. On the theoretical side, multiple studies have proposed methods to characterize entanglement at colliders in terms of various measures in the $h\to WW^*$ and $h\to Z Z^*$ systems\,\cite{Barr:2021zcp,Aguilar-Saavedra:2022wam,Fabbrichesi:2023cev,Fabbri:2023ncz,Ruzi:2024cbt,Wu:2024ovc,Subba:2024aut} primarily in the leptonic decays of the heavy gauge bosons. There are also studies aimed at measuring entanglement in the two-qubit systems such as $t\bar t$ system\,\cite{Fabbrichesi:2021npl,Severi:2021cnj,Aguilar-Saavedra:2022uye,Afik:2022dgh,Dong:2023xiw,Aguilar-Saavedra:2023hss,Aguilar-Saavedra:2024fig,Maltoni:2024csn,Cheng:2024btk} at the LHC and future colliders and in $\tau^-\tau^+$ system\,\cite{Altakach:2022ywa,Ma:2023yvd,Fabbrichesi:2024wcd} at future lepton colliders. On the experimental side, observations and measurements of quantum entanglement in the $t\bar t$ system have been made by the ATLAS and the CMS collaborations in $pp$ collisions at the LHC\,\cite{ATLAS:2023fsd,CMS:2024pts}. These studies are mostly focused on the leptonic decay modes for their cleaner and relatively undistorted signals at the collider detectors. 
   
On the other hand, there has been a growing interest in the muon
collider\,\cite{IMCC,Delahaye:2019omf,InternationalMuonCollider:2024jyv,Accettura:2023ked,Ghosh:2023xbj,Ghosh:2025gdx,Skoufaris:2024okx}, which is able to provide an exceptionally clean environment,
especially for precision Higgs studies\,\cite{Andreetto:2024rra,HiggsMuonCollider2025}. The primary reason behind
the clean environment is the lepton-antilepton nature of initial
states, which is not present in a hadron collider, where the initial state
is composed of quarks and gluons and suffers from significant QCD
backgrounds\,\cite{Han:2021kes,Barik:2024kwv,Das:2024ekt,Abe:2025yur}. Additionally, the energy reach of the muon collider is
higher compared to electron colliders, as there is much less
synchrotron radiation in a muon collider\,\cite{Barklow:2023iav}. Moreover, due to the
absence of energy sharing between partons, the muon collider can deliver
much higher energies for the interactions, with
much less contamination from the hadronic activity\,\cite{AlAli:2021let,Han:2020uid}. Currently, there are
proposals for building a muon collider with energies in 1, 3, and 10 TeV
with integrated luminosities of 10~ab$^{-1}$\,\cite{MuonCollider:2022xlm,Skoufaris:2023jnu,InternationalMuonCollider:2024jyv,InternationalMuonCollider:2025sys}. Therefore, the study
of entanglement in such a machine with high energies would provide crucial
insights into quantum information concepts at the TeV scale.

Our focus in this work is a study of quantum entanglement in the process
$\mu^+ \mu^- \to \nu \bar{\nu} h, h \to ZZ^*$, where the Higgs boson
decays into one on-shell and one off-shell spin-one $Z$ bosons. In the SM,
where the Higgs is a scalar particle, the two $Z$ bosons are produced in
a highly symmetric entangled state, with well-defined helicity
correlations. The joint polarization state of these bosons carries
important information that can be obtained from the angular
distributions of their decay products. The usual way to study the
correlation is via a $9\times9$ density matrix, constructed from the
helicity amplitudes through the leptonic decays of both of the $Z$
bosons\,\cite{Rahaman:2021fcz}.
 
Fully leptonic $ZZ$ decays, while theoretically clean, have low yield
due to the combined branching fraction for both $Z$ bosons decaying to
charged leptons below about 1\%.  The dominant decay mode is when
both $Z$ bosons decay hadronically, which is about 50\%, but this mode is
messy due to their hadronic nature. However, in the mode where one $Z$ boson decays
leptonically and the other decays hadronically, producing two
visible jets in the detector along with two leptons, provides a relatively larger
event yield in a relatively clean environment. Although it 
introduces a level of experimental complications, the greater yield helps in achieving a better-defined signal region. The usual notion is
that angular reconstruction of hadronic $Z$ decays is inherently less
precise due to jet clustering ambiguities, hadronization, and detector
effects. These factors can smear or
distort the angular variables needed to reconstruct the density matrix, thereby
reducing sensitivity to entanglement and nonlocality signals. With the advancement of techniques to tackle final states with jets, specifically in the study of spin polarization\,\cite{De:2020iwq,Dey:2021sug,De:2024puh,Ghosh:2025gue}, we attempt to study final states with hadronic decays of the $Z$ boson. A central
goal of this work is to determine how well entanglement can still be
measured in the presence of these limitations.
 
In this paper, we present a comprehensive study of quantum entanglement
and Bell's inequality violation in the leptonic-hadronic decay channel of
the $h\to Z Z^*$ process. We have performed a thorough study, whose key points may be summarized as follows.
\begin{itemize}
\item We go beyond the fully leptonic channels and perform an analysis in leptonic and hadronic decay modes to study entanglement at a muon
      collider with detector effects included using Delphes. 
\item With the help of optimized cuts on the phase space, we select a signal-enriched region to improve the quantification of
      entanglement correlations.
\item Furthermore, we use unfolding techniques to further
      improve the entanglement measurements in terms of Bell inequality
      violation.
\end{itemize}
  
The remainder of this paper is organized as follows. In
Section~\ref{sec:theory}, we introduce the formalism for describing the
$ZZ^*$ spin density matrix. We then describe the construction of the angular observables from the decay products of $Z$ bosons in \ref{sec:ang-dist}. Section~\ref{sec:evt-gen} describes the Monte Carlo
event generation, including detector simulation. In
Section~\ref{sec:AC_lljj}, we outline the methods used to
reconstruct spin correlations, and we outline the method of unfolding to reconstruct these distributions at the truth-level in section~\ref{sec:unfold}.
Our results are presented and
discussed in Section~\ref{sec:results}, and we conclude in
Section~\ref{sec:conclusion} with a summary and outlook for future studies.

\section{Theoretical Framework}\label{sec:theory}
\subsection{Density Matrix and Bell Nonlocality}\label{sec:bell-nonlocality}
For a quantum system with a statistical mixture of a collection of
normalized states $\{\ket{\psi_i}\}$ with weights $\{p_i\ge0\}$ and
$\sum_i p_i = 1$ is described by the density matrix
\begin{eqnarray}
\rho = \sum_i p_i \ket{\psi_i} \bra{\psi_i}.
\end{eqnarray}
By construction, the density matrix is a linear operator that can act on
any states in the Hilbert space describing the system. A few basic
properties of the density matrix are (a) hermiticity:
$\rho^\dagger = \rho$, (b) normalization: Tr[$\rho$] = 1, and (c) positive
definiteness: $\expval{\rho}{\psi} \ge 0$ for any $\ket{\psi}$ belonging
to the Hilbert space. With the density matrix, the expectation value of an observable $\operatorO$ can be expressed using the formula
$\expectationO = \Tr[\rho \operatorO]$.

In the case of a bipartite system, {\it i.e.}, a system $S$ consisting of two
subsystems $S_A$ and $S_B$, the system can be described by the tensor
product space of the two Hilbert spaces (say $\mathcal{H}_A$ and
$\mathcal{H}_B$) describing the systems $S_A$ and $S_B$. If one chooses
bases for the vector spaces $\mathcal{H}_A$ and $\mathcal{H}_B$ to be
$\{\ket{\alpha}_A\}$ and $\{\ket{\beta}_B\}$, respectively, the system $S$
can be described in the basis $\{\ket{\alpha}_A\otimes\ket{\beta}_B\}$.
Therefore, a state $\ket{\psi}$ of $S$ can be expressed as
$\ket{\psi} = \sum_{\alpha,\beta} \psi_{\alpha\beta} \ket{\alpha}_A \otimes \ket{\beta}_B$.
Apart from the procedural complication due to the several indices, the
constructions and properties of the density matrix for a statistical admixture of a bipartite system follow the same principles as described above. 

Quantum entanglement in terms of a bipartite system can be stated in
terms of the density matrix. The two subsystems are said to be separable
if the density matrix describing the system can be written as a convex
sum of products of density matrices of the individual subsystems, {\it i.e.},
\begin{eqnarray}
\rho_S = \sum_{i} p_i\,\rho_{A,i} \otimes \rho_{B,i}, \qquad {\rm with }\quad p_i \ge 0 \quad{\rm and } \sum_i p_i = 1.\label{eq:sep-def}
\end{eqnarray}
Here, $\rho_A$ and $\rho_B$ are density operators defined in the Hilbert
spaces $\mathcal{H}_A$ and $\mathcal{H}_B$, respectively. The system
described by $\rho_S$ is called entangled if it is not separable.
Although the definition for separable states looks simple, there is no known general method to determine whether a system is entangled for an arbitrary $\rho_S$ in an arbitrary dimension.
Only for two-qubit and qubit-qutrit systems can general tests be
applied with the help of Peres-Horodecki
criterion\,\cite{Peres:1996dw,Horodecki:1997vt}. 
There are a few different formalisms of determining the degree of entanglement using measures such as von Neumann entropy\,\cite{Cerf:1995sa,Nielsen:2012yss}, concurrence\,\cite{Bennett:1996gf,Wootters:1997id,Rungta:2001zcj}, violation of Bell's inequality\,\cite{Bell:1964kc,Brunner:2013est,Scarani:2019zug}. We work in the formalism of Bell's inequality violation\,\cite{Bell:1964kc}. In this framework, entanglement is a necessary but not sufficient condition for Bell nonlocality. In other words, a system may be entangled without violating Bell's inequality. 

For classical and locally realizable systems, Bell's inequalities provide a powerful criterion, whose violation strongly suggests the non-local and quantum behaviour. For the two-qutrit system, Bell's inequality can be formulated as follows. Two observers, A and B, each perform two independent measurements for observables $\qty(A_1, A_2)$ and $\qty(B_1, B_2)$ defined in $\mathcal{H}_A$ and $\mathcal{H}_B$, respectively. Each observable has three possible outcomes: $\pm 1$ and 0. One then constructs the quantity 
\begin{eqnarray}
\Ithree &=& P(A_1 = B_1) + P(B_1 = A_2 + 1) + P(A_2 = B_2) + P(B_2 = A_1) \nonumber\\
&&- \left[P(A_1 = B_1 - 1) + P(B_1 = A_2) + P(A_2 = B_2 - 1) + P(B_2 = A_1 - 1)\right],
\label{eq:i3}
\end{eqnarray}
where $P(B_i = A_j + m)$ is the probability that the outcomes for $B_i$ and $A_j$ differ by $m$ modulo 3. For classical local systems, the value of $\Ithree$ should never exceeds 2, {\it i.e.}, $\Ithree \leq 2$. A violation of this inequality indicates the presence of nonlocality and entanglement. 

Now the probabilities in Eq.~(\ref{eq:i3}) can be expressed in terms of expectation values of suitable projection operators. Thus $\Ithree$ can be estimated from the density matrix using an appropriate operator, known as the Bell operator. In terms of a carefully constructed Bell operator $\bellOperator$, the Bell inequality becomes
\begin{eqnarray}
\Ithree = \expval{\bellOperator} = \Tr[\rho\bellOperator] \le 2. \label{eq:I3}
\end{eqnarray}

This formalism is analogous to the CHSH inequality\,\cite{Clauser:1969ny}, which is optimal for a two-qubit system. The present formalism, called CGLMP inequality\,\cite{Collins:2001qdi}, is the general version suitable for two qudit systems of arbitrary Hilbert space dimension. 
 
\subsection{Spin Density Matrix for $h\to ZZ^*$ System}
\label{sec:density}
We want to study the Bell's inequality violation in the process $\mu^+ \mu^- \to \nu\bar{\nu} h; h\to ZZ^*$. Specifically, we want to concentrate on the part $h\to Z Z^*$, {\it i.e.}, the production of one on-shell $Z$ boson and one off-shell $Z$ boson, which is represented by $Z^*$. Therefore, all the extra indices and momenta of the neutrino and anti-neutrino pair are summed over or integrated out appropriately, keeping only the helicity indices of the $ZZ^*$ system. Since the $ZZ^*$ system arises from the decay of the scalar $h$, the presence of a neutrino and anti-neutrino pair should not affect the spin-correlation of the $ZZ^*$ system. In any case, the helicity states of any $ZZ^*$ system can be described by a $9\times 9$ density operator\,\cite{Rahaman:2021fcz} acting on a 9 dimensional Hilbert space arising from $3\otimes 3$ helicity states corresponding to 3 spin states of each $Z$ boson. The joint density matrix can be written in the helicity basis as
\begin{equation}
	\rho_{\lambda_1 \lambda_2, \lambda_1' \lambda_2'} = \mathcal{N}_\rho \mathcal{M}_{\lambda_1 \lambda_2} \, \mathcal{M}^*_{\lambda_1' \lambda_2'},
	\label{eq:densitymatrix_helicity}
\end{equation}
where $\mathcal{M}_{\lambda_1 \lambda_2}$ is the helicity amplitude for $h \to Z(\lambda_1) Z^*(\lambda_2)$, and $\mathcal{N}_\rho$ is a normalization factor to ensure $\Tr[\rho] = 1$. The helicity amplitudes are determined by the interaction Lagrangian of the Higgs boson with the $Z$ bosons. In the Standard Model (SM), the scalar nature of the Higgs implies that only the combinations with $\lambda_1 = -\lambda_2$ are allowed, leading to a highly entangled state in the helicity basis.

Choosing the $z$-axis along the direction of the $Z$ boson momentum, we take the eigenstates of the spin projection operator $J_z$ as the basis vectors of the spin space.
The matrix form of the density matrix can be constructed using a basis of the tensor product space of the two helicity spaces corresponding to $Z$ and $Z^*$. In other words, the basis for the bipartite system is $\left\{\ket{-,-}, \ket{-,0}, \ket{-,+}, \ket{0,-}, \ket{0,0}, \ket{0,+}, \ket{+,-}, \ket{+,0}, \ket{+,+}\right\}$, where the first entry comes from the $Z$ and the second comes from $Z^*$.

We will be following the computation as described in Ref.\,\cite{Aguilar-Saavedra:2022wam}. We outline only the key elements of the construction for completeness, and refer to Ref.\,\cite{Aguilar-Saavedra:2022wam} for further details. An important subtlety is that the two $Z$ bosons are not simultaneously on-shell, since the mass of the Higgs boson is less than $M_Z$. Typically, one boson is on-shell while the other is off-shell. The off-shell $Z$ boson propagator has a scalar component, which vanishes for decays into massless fermion pairs. Furthermore, the off-shell $Z$ boson mass is distributed with a probability density, which must be properly incorporated. After accounting for these effects, the density matrix takes the general form, calculated in the $ZZ^*$ rest frame,  
\begin{eqnarray}
\rho = \frac{1}{2+\chi^2}\begin{pmatrix}
0 & 0 & 0   & 0 & 0      & 0 & 0   & 0 & 0 \\
0 & 0 & 0   & 0 & 0      & 0 & 0   & 0 & 0 \\
0 & 0 & 1   & 0 & \xi    & 0 & 1   & 0 & 0 \\ 
0 & 0 & 0   & 0 & 0      & 0 & 0   & 0 & 0 \\
0 & 0 & \xi & 0 & \chi^2 & 0 & \xi & 0 & 0 \\ 
0 & 0 & 0   & 0 & 0      & 0 & 0   & 0 & 0 \\
0 & 0 & 1   & 0 & \xi    & 0 & 1   & 0 & 0 \\ 
0 & 0 & 0   & 0 & 0      & 0 & 0   & 0 & 0 \\
0 & 0 & 0   & 0 & 0      & 0 & 0   & 0 & 0 \\
\end{pmatrix},
\label{eqn:density-ZZ}
\end{eqnarray}
where $\xi$ and $\chi$ encode off-shell $Z$ boson effects. One should remember that this construction is in the rest frame of the $ZZ^*$ system. 
This specific form of the density matrix of the $ZZ^*$ system can be understood from the fact that $Z$ and $Z^*$ originate from the spin-zero Higgs boson decay. Therefore, the $ZZ^*$ state is restricted to three possible joint spin configurations belonging to $\{\ket{-,+}, \ket{0,0}, \ket{+,-}\}$. Thus, only the 9 components of the full density matrix survive at the $i$ and $\jth$ position with $i,j \in \qty{3,5,7}$. 

There are various useful parametrizations available for a $9\times 9$ spin density operator arising from two spin-one states. In this work, we use the tensor product basis $T^L_M$ with $L=0,1,2$ and $M=0, \cdots, \pm L$. Therefore the basis for the $9\times 9$ density matrix is $\qty{T^L_M\otimes T^{L'}_{M'}}$, with each of $T^L_M$ and $T^{L'}_{M'}$ belonging to $\qty{T_0^0, T^1_0, T^1_{\pm 1}, T^2_0, T^2_{\pm 1}, T^2_{\pm 2}}$. The explicit forms for these are
\begin{eqnarray}
&&T^1_0 = \sqrt{\frac{3}{2}} \begin{pmatrix} 1 & 0 & 0 \\ 0 & 0 & 0 \\ 0 & 0 & -1\end{pmatrix}, \qquad
T^1_{-1} = \sqrt{\frac{3}{2}} \begin{pmatrix}  0 & 0 & 0 \\ 1 & 0 & 0 \\ 0 & 1 & 0\end{pmatrix}, \qquad
T^1_1 = \sqrt{\frac{3}{2}} \begin{pmatrix} 0 & -1 & 0 \\ 0 & 0 & -1 \\ 0 & 0 & 0\end{pmatrix},\\
&& T^0_0 = \mathbb{I}_3, \qquad\qquad\qquad \qquad T^2_0 = \frac{\sqrt{2}}{3}\qty(T^1_1T^1_{-1} + T^1_{-1}T^1_1 + 2T^1_0T^1_1), \\
&& T^2_{\pm 1} = \sqrt{\frac{2}{3}} \qty(T^1_{\pm 1} T^1_0 + T^1_0 T^1_{\pm 1}), \qquad\qquad\qquad\quad T^2_{\pm 2} = \frac{2}{\sqrt 3} T^1_{\pm 1}T^1_{\pm 1}.
\end{eqnarray}
These operators satisfy the normalization condition $\Tr[T^L_M\qty(T^{L'}_{M'})^\dagger] = 3 \delta_{LL'} \delta_{MM'}$, and they are traceless except for $T^0_0$. With this choice of basis, the density matrix $\rho$ can be expressed as\,\cite{Aguilar-Saavedra:2015yza, Aguilar-Saavedra:2017zkn,Rahaman:2021fcz}
\begin{equation}
	\rho = \frac{1}{9} \left[
	\mathbb{I}_3 \otimes \mathbb{I}_3
	+ A_{LM}^1 T_M^L \otimes \mathbb{I}_3
	+ A_{LM}^2 \mathbb{I}_3 \otimes T_M^L
	+ C_{L M, L' M'} T_M^L \otimes T_{M'}^{L'}
	\right].
	\label{eq:rho2}
\end{equation}
The coefficients $A_{LM}^i$ describe single-boson polarizations, while $C_{L M, L' M'}$ encode spin correlations between the two $Z$ bosons. In Eq.~(\ref{eq:rho2}), implicit sums over the indices $L, L', M$, and $M'$ are assumed with $L,L'=1,2$,  $-L\le M\le L$, and $-L'\le M' \le L'$. 
Furthermore, the $A$ and $C$ coefficients satisfy
\begin{eqnarray}
	A^{1,2}_{L,M} =(-1)^M \qty(A^{1,2}_{L,-M})^*, \qquad\qquad C_{L M, L' M'} = (-1)^{M+M'} \qty(C_{L,-M,L',-M'}).
\end{eqnarray}
Using these relations and the form of the density matrix in Eq.~(\ref{eqn:density-ZZ}), the nonzero components are fully determined in terms of the coefficients $A^1_{2,0}$, $C_{2,1,2,-1}$, and $C_{2,2,2,-2}$. The explicit form of the density matrix is given by
\begin{eqnarray}
	\rho = \begin{pmatrix}
		0 & 0 & 0 & 0 & 0 & 0 & 0 & 0 & 0 \\
		0 & 0 & 0 & 0 & 0 & 0 & 0 & 0 & 0 \\
		0 & 0 & \frac{2+\sqrt{2} A^1_{2,0}}{6} & 0 & \frac{C_{2,1,2,-1}}{3} & 0 & \frac{C_{2,2,2,-2}}{3} & 0 & 0\\ 
		0 & 0 & 0 & 0 & 0 & 0 & 0 & 0 & 0 \\
		0 & 0 & \frac{C_{2,1,2,-1}}{3} & 0 & \frac{1 - \sqrt{2} A^1_{2,0}}{3} & 0 &  \frac{C_{2,1,2,-1}}{3} & 0 & 0\\ 
		0 & 0 & 0 & 0 & 0 & 0 & 0 & 0 & 0 \\
		0 & 0 & \frac{C_{2,2,2,-2}}{3} & 0 & \frac{C_{2,1,2,-1}}{3} & 0 & \frac{2+\sqrt{2} A^1_{2,0}}{6} & 0 & 0\\ 
		0 & 0 & 0 & 0 & 0 & 0 & 0 & 0 & 0 \\
		0 & 0 & 0 & 0 & 0 & 0 & 0 & 0 & 0 \\
	\end{pmatrix}.\label{eq:rho3}
\end{eqnarray}

Moreover, following Ref.\,\cite{Aguilar-Saavedra:2022wam}, the Bell operator that maximally captures\,\cite{Collins:2001qdi} the Bell's inequality in the $h\to ZZ^*$ system can be written as
\begin{align}
	\bellOperator =& \left[\frac{2}{3\sqrt{3}}\left( T_{1}^{1}\otimes T_{1}^{1}-T_{0}^{1}\otimes T_{0}^{1}+T_{1}^{1}\otimes T_{-1}^{1} \right)
	+\frac{1}{12}\left( T_{2}^{2}\otimes T_{2}^{2}+T_{2}^{2}\otimes T_{-2}^{2} \right)\right. \nonumber\\
	&\left.+\frac{1}{2\sqrt{6}}\left(T_2^2\otimes T_0^2+T_0^2\otimes T_2^2\right)
	-\frac{1}{3}(T_1^2\otimes T_1^2+T_1^2\otimes T_{-1}^2)
	+\frac{1}{4}T_0^2\otimes T_0^2 \right] + \mathrm{h.c.} \label{eq:bell}
\end{align}

Finally, using Eq.~(\ref{eq:I3}) and expressions for $\rho$ [Eq.~(\ref{eq:rho3})] and $\bellOperator$ [Eq.~(\ref{eq:bell})], the expectation value of the Bell operator takes the from
\begin{equation}
	\Ithree = \frac{1}{36}\left( 18 + 16\sqrt{3}
	- \sqrt{2}(9 - 8\sqrt{3})A_{2,0}^{1}
	- 8(3 + 2\sqrt{3})C_{2,1,2,-1}
	+ 6C_{2,2,2,-2} \right).
	\label{eq:I_3}
\end{equation}

The advantage of this parametrization is that it allows us to extract the values of these $A$ and $C$ coefficients from the angular variables of the decay products of the $Z$ and $Z^*$ bosons. Reconstruction of these observables is the discussion of the next subsection.

\subsection{Angular Distributions and Density Matrix Reconstruction from the $Z$ Bosons Decay}\label{sec:ang-dist}

The decay density matrix for the decay of a $Z^{(*)}$ boson to a pair fermion and anti-fermion pair $f\bar f$ can be expressed as
\begin{equation}
\Gamma_{f,\lambda\lambda'} = \mathcal{N}_\Gamma\mathcal{A}_\lambda \mathcal{A}_{\lambda'}^*,
\end{equation}
where $\mathcal{A}_\lambda$ denotes the decay amplitude for $Z^{(*)} (\lambda)\to f \bar{f}$ with helicity $\lambda$, and $\mathcal{N}_\Gamma$ is a normalization coefficient. In the rest frame of the $Z^{(*)}$ boson with the choice of the $z$-axis chosen as the propagation direction of the decaying boson, the decay matrix takes the form\,\cite{Aguilar-Saavedra:2022wam,Aguilar-Saavedra:2022mpg}
\begin{equation}
	\Gamma_f(\theta, \phi) = \frac{1}{4}
	\begin{pmatrix}
		1 + \cos^2 \theta - 2 \eta_f \cos \theta & \sqrt{2} (\cos\theta - \eta_f) \sin \theta^{i\phi} & \sin^2 \theta e^{i2\phi} \\
		\sqrt{2} (\cos\theta - \eta_f) \sin\theta e^{-i\phi} & 2 \sin^2 \theta & -\sqrt{2} (\cos\theta + 2\eta_f) \sin\theta e^{i\phi} \\
		\sin^2 \theta\,e^{-i2\phi} & -\sqrt{2} (\cos \theta + \eta_f) \sin \theta e^{-i\phi} & 1 + \cos^2 \theta + 2 \eta_f \cos \theta
	\end{pmatrix}. \label{eq:decay-density}
\end{equation}
Here, $\theta$ and $\phi$ are the polar and azimuthal angles of the fermion $f$ in the $Z^{(*)}$ rest frame. The decay density matrix with respect to anti-fermion $\bar{f}$ satisfies $\Gamma_{\bar{f}} (\theta,\phi) = \Gamma_f(\pi-\theta,\pi+\phi)$. In Eq.~(\ref{eq:decay-density}), $\eta_f$ is the asymmetry parameter of the left- and right-chiral couplings of the fermion to the $Z^{(*)}$ boson:
\begin{eqnarray}
\eta_f = \frac{C_L^2 - C_R^2}{C_L^2 + C_R^2} = \frac{1-4|Q_f|\,s^2_w}{1-4|Q_f|\,s^2_w+8Q_f^2\,s^4_w}, 
\end{eqnarray}
where $C_L$ and $C_R$ are the left- and right-chiral couplings, and $s_w$ is the sine of the weak mixing angle. In the SM, the values of $\eta_f$ are
\begin{eqnarray}
\begin{matrix}
\eta_\nu = 1, &\qquad \eta_\ell = 0.13, & \qquad \eta_u = 0.67, & \qquad {\rm and } & \eta_d = 0.94. 
\end{matrix}
\end{eqnarray}

In terms of the density matrix of the $ZZ^*$ system, the decay density matrices of $Z\to f_1\bar{f_1}$ and $Z^*\to f_2\bar{f_2}$, the differential cross section can be expressed as\,\cite{Rahaman:2021fcz,Aguilar-Saavedra:2022wam}
\begin{eqnarray}
\frac{1}{\sigma} \frac{d\sigma}{d\Omega_1 \, d\Omega_2}
    &=& \left(\frac{3}{4\pi}\right)^2 \sum_{\lambda_1,\lambda_2,\lambda_1',\lambda_2'} \rho_{\lambda_1\lambda_2,\lambda_1'\lambda_2'} \Gamma_{f_1, \lambda_1\lambda_1'} \Gamma_{f_2, \lambda_2\lambda_2'}\\
    &=& \left(\frac{3}{4\pi}\right)^2
    \Tr[\rho \qty(\Gamma_{f_1} \otimes \Gamma_{f_2}^T)],
    \label{eq:xsection}
\end{eqnarray}
where $d\Omega_1$ and $d\Omega_2$ are the solid angle elements for $f_1$ and $f_2$ in the respective $Z$ boson rest frames. Furthermore,
\begin{eqnarray}
	&& \Tr[T^0_0 \Gamma^T_f] = 2\sqrt{\pi}\,Y^0_0(\theta,\phi), \\
	&& \Tr[T^1_M \Gamma^T_f] = -\sqrt{2\pi}\,\eta_f\,Y^M_1(\theta,\phi), \\
	&& \Tr[T^2_M \Gamma^T_f] = \sqrt{\frac{2\pi}{5}}\,Y^M_2(\theta,\phi), \qquad
\end{eqnarray}
where $Y^M_L$ are the spherical harmonics. Equation~(\ref{eq:xsection}) can be recast as
\begin{eqnarray}
	\frac{1}{\sigma}\frac{d\sigma}{d\Omega_1 d\Omega_2}
	= \frac{1}{(4\pi)^2}
	\bigg[
	1 &&+ A^1_{LM} Y_L^M(\Omega_1)
	+ A^2_{LM} Y_L^M(\Omega_2)  \\
	 &&+ \sum_{L, M, L', M'} C_{LM, L'M'}
	Y_L^M(\Omega_1)Y_{L'}^{M'}(\Omega_2)
	\bigg].
\end{eqnarray}
Using the orthogonality of the spherical harmonics, the angular coefficients can be written as
\begin{eqnarray}
    &&\int \frac{1}{\sigma}\frac{d\sigma}{d\Omega_i}Y_L^M(\Omega_i)^*\, d\Omega_i = \frac{B_L}{4\pi}A_{LM}^i, \\ \label{eq:ACLM1}
    &&\int \frac{1}{\sigma}\frac{d^2\sigma}{d\Omega_1\,d\Omega_2}\,Y_L^M(\Omega_1)^*\,Y_{L'}^{M'}(\Omega_2)^*\,d\Omega_1\,d\Omega_2 = \frac{B_L B_{L'}}{(4\pi)^2}C_{LM, L'M'},
    \label{eq:ACLM}
\end{eqnarray}
with $B_1 = -\sqrt{2\pi}\,\eta_f$, $B_2 = \sqrt{2\pi/5}$.

Once the coefficients $A_{LM}^i$ and $C_{LM, L'M'}$ are obtained, the value of $\Ithree$ follows directly from Eq.~(\ref{eq:I_3}). Importantly, the estimation of $\Ithree$ does not involve any $A$ or $C$ coefficients with $L=1$; hence, it does not depend on the specific fermion mode of the $Z$ boson decay.
In the computing $\Ithree$, we use fermions as quarks or negatively charged leptons for the decay density matrix in Eq.~(\ref{eq:decay-density}) in the $Z^{(*)}$ rest frame. If instead antiquarks or positively charged leptons are used, the decay density matrix can be obtained simply by transforming $\theta \to \pi - \theta$ and $\phi \to \pi +\phi$, which implies the transformation $Y_L^M(\theta,\phi) \to Y_L^M(\pi - \theta, \pi +\phi) = (-1)^LY_L^M(\theta,\phi)$. Since only $L=2$ coefficients enter into $\Ithree$, its value is independent of the choice of fermion or its antiparticle.

We also note that the index $i=1$ in the solid angle measure $d\Omega$ and $A_{LM}$ corresponds to the on-shell $Z$-boson, and $i=2$ is for the off-shell one. In what follows, we denote the on-shell and off-shell bosons as $Z_1$ and $Z_2$, respectively, and also use the notation $Z$ and $Z^*$ interchangeably to refer to the same pair.

\begin{figure}[h]
	\centering 
	\subfloat[]{\label{Fig:Z3-relicupper}\includegraphics[width=0.50\textwidth]{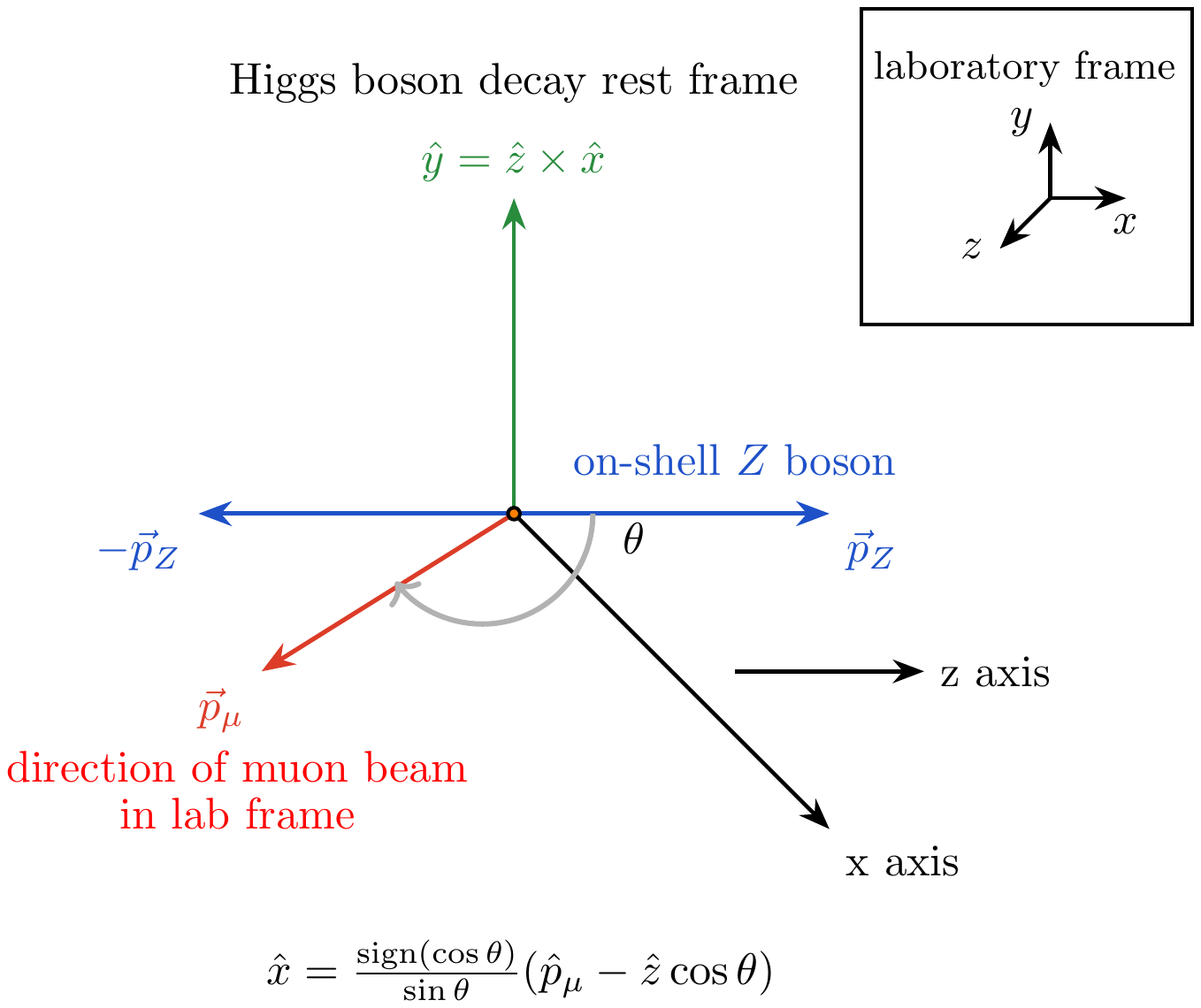}} ~
	\subfloat[]{\label{Fig:Z3-relicupperlower}\includegraphics[width=0.50\textwidth]{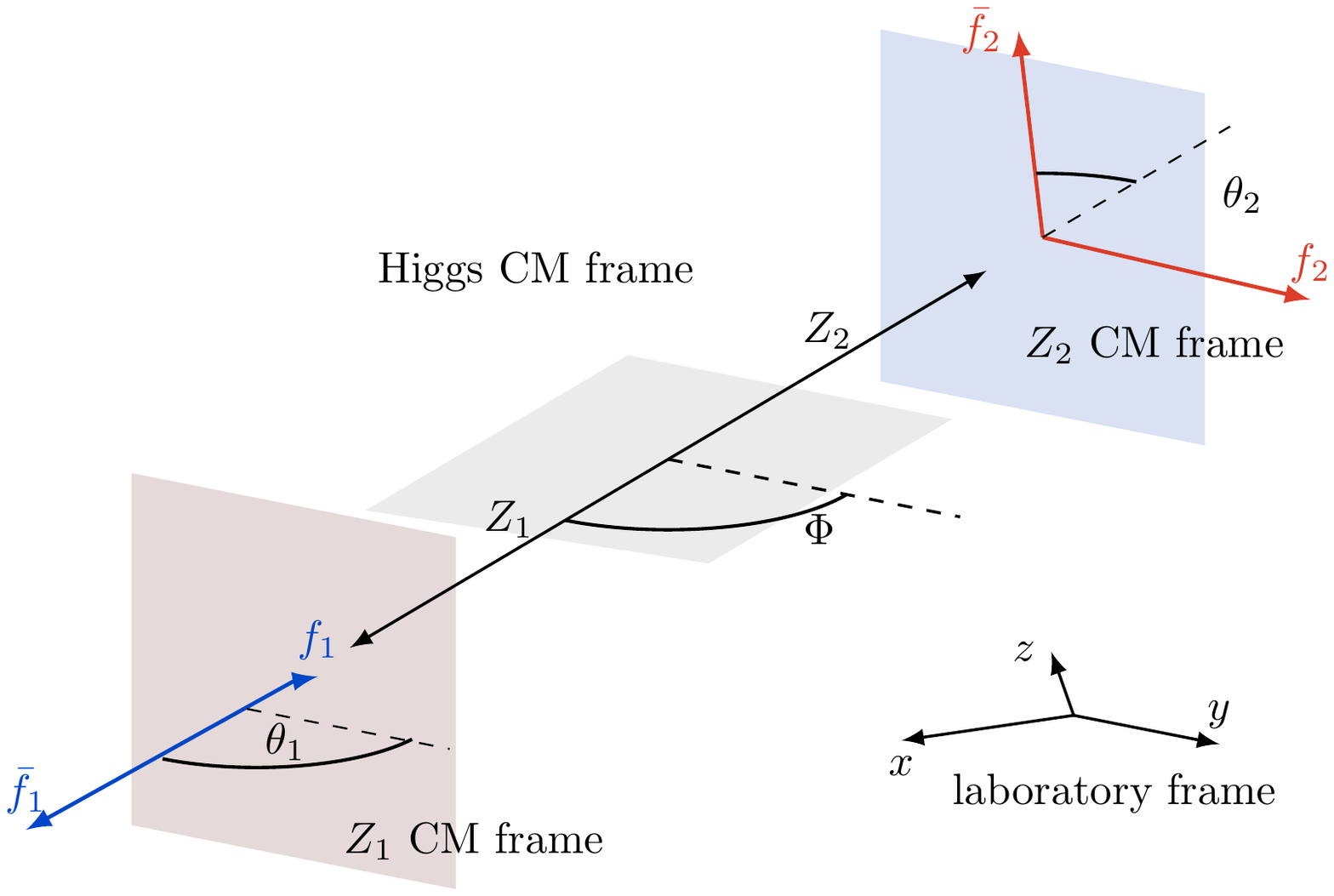}} \\
	\caption{Definition of reference frames. In the Higgs CM frame, the direction of the on-shell $Z$ boson defines the positive $z$-axis (left). The same coordinate system is employed to describe the angular distributions of the final-state fermion and anti-fermion pairs in their respective $Z$ rest frames (right).}
	\label{fig:frame}
\end{figure}
Based on the developments above, without loss of generality, we perform our analysis using the reference frames and coordinate system illustrated in Fig.~\ref{fig:frame}.  
The spin eigenstates of each $Z$ boson are defined in their own rest frames, which are boosted into the Higgs rest frame for consistency.  
Following the convention used in Refs.\,\cite{Aguilar-Saavedra:2022wam,Fabbrichesi:2023cev,Barr:2024djo}, we adopt:
\begin{itemize}
	\item $\hat{z}$: the unit vector in the direction of the on-shell $Z$ boson ($Z_1$) three-momentum in the Higgs rest frame.
	\item $\hat{x}$: the unit vector in the production plane, defined as $\hat{x} = {\rm sign}(\cos\theta)\,(\hat{p}_\mu - \hat{z}\cos\theta)/\sin\theta$, where $\hat{p}_\mu = (0, 0, 1)$ represents the beam direction in the lab frame.
	\item $\hat{y} = \hat{z} \times \hat{x}$: the unit vector orthogonal to the production plane.
\end{itemize}
After fixing the reference frame as the Higgs boson rest frame and the above coordinate system, the solid angles $\Omega_1$ and $\Omega_2$ are subsequently computed in the respective rest frames of $Z_1$ and $Z_2$ by boosting back from the Higgs rest frame.

\section{Monte Carlo Event Samples and Angular Variables Construction}\label{sec:simulation}
In this section, we describe the Monte Carlo event generation and detector simulation framework employed in our analysis. We also outline the construction of useful variables from the reconstructed leptons and jets. We then discuss the selection criteria applied to those variables to isolate the signal-enriched region.

\subsection{Event Generation}\label{sec:evt-gen}
The simulation of the Born-level signal process
\begin{equation}
\mu^-\mu^+ \to \nu \bar\nu h \to \nu\bar\nu ZZ^* \to \nu\bar\nu  \ell^+\ell^- j j
\end{equation}
at a muon collider is performed using \texttt{MadGraph5\_aMC@NLO}\,\cite{Alwall:2011uj}. For the leptons, we consider $\ell^\pm=e^\pm,\mu^\pm,\tau^\pm$ and for quarks, we include all five quark flavours $u,d,s,c,b$. The events are generated at centre-of-mass energies $\sqrt{s} = 1$, 3, and 10 TeV corresponding to an integrated luminosity of $\lumi = 10~\text{ab}^{-1}$, representative of the projected operational scenarios of future muon colliders\,\cite{InternationalMuonCollider:2025sys}. 

The background contributions considered in this analysis include the following dominant SM processes:
\begin{center}
\begin{tabular}{rl}
$\boldsymbol{\ell\ell j j}$: & $\mu^+ \mu^- \rightarrow \ell^+ \ell^- j j$ \\
$\boldsymbol{WWZ}$: &$\mu^+ \mu^- \rightarrow W^+(\ell^+\nu) W^-(\ell^-\bar\nu) Z(jj)$ \\
$\boldsymbol{t \bar t}$: & $\mu^+ \mu^- \rightarrow t \bar{t}$\\
\end{tabular}
\end{center}
Here again, $\ell^\pm=e^\pm,\mu^\pm,\tau^\pm$, and $j$ denote all five quarks or anti-quarks. These channels collectively encompass backgrounds that can mimic the signal topology. For the $\ell\ell jj$ background sample, the diagrams with the Higgs boson in the intermediate state are excluded as they produce the signal topology. For the $t\bar t$ background, fully leptonic, i.e., both tops decay leptonically, and semileptonic, i.e., one decays hadronically and the other leptonically, modes are considered.
During the generation of parton-level events for the signal and background processes using \texttt{MadGraph5\_aMC@NLO}, the following basic kinematic cuts are imposed uniformly on signal and background events.
\begin{eqnarray}
	{p_T}_j > 5.0~{\rm GeV}, \quad {p_T}_\ell > 5.0~{\rm GeV}, \quad |\eta_j| < 5.0,\quad {\rm and}~|\eta_\ell| < 3.0 
\end{eqnarray}
The expected numbers of signal and background events at the projected luminosity have been estimated and are summarized in Table~\ref{events}.
\begin{table}[htbp]
\centering
\begin{tabular}{|l|r|r|r|r|r|r|}
\hline
\multirow{2}{*}{Process} & \multicolumn{2}{|c|}{1 TeV} & \multicolumn{2}{|c|}{3 TeV} & \multicolumn{2}{|c|}{10 TeV} \\
\cline{2-7}
 & $\mathbf{\sigma}$ [fb] & $N_{\rm evt}$~~\, & $\mathbf{\sigma}$ [fb] & $N_{\rm evt}$\,~~ & $\mathbf{\sigma}$ [fb] & $N_{\rm evt}$~~ \\
\hline
Signal        &   0.433 &      4\,330 &  0.961 &   9\,610 & 1.35  & 13\,500 \\
\hline
$\ell\ell jj$ & 144.0   & 1\,440\,000 & 48.0   & 480\,000 & 9.21  & 92\,100 \\
\hline
$t\bar{t}$    &  82.0   &    820\,000 &  9.44  &  94\,400 & 0.865 &  8\,650 \\
\hline
$WWZ$         &   3.85  &     38\,500 &  1.68  &  16\,800 & 0.387 &  3\,870 \\
\hline
\end{tabular}
\caption{Cross sections and expected number of events ($N_{\rm evt}$) at $\lumi = $ 10 ab$^{-1}$.}
\label{events}
\end{table}
The generated parton-level events are subsequently processed through \texttt{Pythia8}\,\cite{Bierlich:2022pfr} for parton showering and hadronization.
For detector effects on the final state leptons and hadrons, we use {\tt Delphes}\,\cite{deFavereau:2013fsa} fast detector simulation with the default Delphes card for muon collider\,\cite{delphes_mu_coll_card,delphes_mu_coll_card_git}.  For this study, jets are reconstructed using the anti-$k_T$ algorithm\,\cite{Cacciari:2008gp} with a typical jet radius $R = 0.5$. 

\subsection{Computation of $A_{LM}$ and $C_{LM, L'M'}$ Coefficients}\label{sec:AC_lljj}
We can see from Eq.~(\ref{eq:I_3}) that, after narrowing down all the calculations, the only task remaining is to calculate the values of $A$ and $C$ coefficients. As described, the angular coefficients $A_{LM}^i$ and $C_{LM, L'M'}$ characterize the spin and polarization structure of the intermediate $ZZ^{(*)}$ system in the decay $h \to ZZ^{(*)} \to f_1 \bar{f_1} f_2\bar{f_2}$. These coefficients can be extracted from the events with the help of Eqs.~(\ref{eq:ACLM1}) and (\ref{eq:ACLM}) after choosing the coordinate system described in section~\ref{sec:ang-dist} via Fig.~\ref{fig:frame}. Thus, given any event sample, the computation of the $A$ and $C$ coefficients can be expressed in terms of the expectation values of the spherical harmonics as follows:
\begin{eqnarray}
A_{LM}^i     = \frac{B_L}{4\pi} \expval{\tilde{A}_{LM}^i}  \qquad \qquad 
C_{LM, L'M'} = \frac{B_L B_{L'}}{\qty(4\pi)^2} \expval{\tilde{C}_{LM, L'M'}},
\end{eqnarray}
with the expressions for the variables given as
\begin{eqnarray}
\tilde{A}_{LM}^i =  Y^M_L\qty(\Omega_i)^* \qquad \qquad \tilde{C}_{LM, L'M'} =   Y^M_L\qty(\Omega_i)^*\, Y^{M'}_{L'}\qty(\Omega_i)^*. \label{eq:ACtilde}
\end{eqnarray}

In this work, we have particularly focused on the final state of the process $h\to ZZ^*$, where one $Z$ boson decays hadronically and the other decays leptonically. In the hadronic mode of $Z$ boson, the quark and antiquark pair, after showering and hadronization, subsequently form jets. We then used the following methods to compute the $A$ and $C$ coefficients.
\begin{itemize}
\item After reconstruction, each event must contain two opposite-sign, same-flavor leptons and at least two jets.

\item We then construct two candidate $Z$ bosons, one from the two reconstructed leptons and one from the two leading jets. Here, `leading' refers to the object with the highest $p_T$.

\item The candidate $Z$ boson with invariant mass closest to $M_Z$ is assigned the on-shell $Z$ boson ($Z_1$) and the other is assigned the off-shell $Z$ boson ($Z_2$). We also use the notation $Z$ and $Z^*$ interchangeably for this pair.

\item The quantities $\tilde{A}^i_{LM}$ and $\tilde{C}_{LM,L'M'}$ are computed in each event using expressions in Eq.~(\ref{eq:ACtilde}) and the coordinate system defined in Fig.~\ref{fig:frame}.
\end{itemize}

In the hadronic decay mode of the on-shell or the off-shell $Z$ boson, the produced quarks, which capture the spin correlation, cannot be reconstructed accurately. This is because the quarks are not detected in isolation; rather, they are captured as jets after the hadronization process, which smears the momentum information of the original quark or anti-quark. They are further smeared by the detector effects. We therefore use the final reconstructed jets as proxies for the quarks or anti-quarks. Consequently, difficulty arises in the reconstruction of the hadronic objects, such as jets. As a result, some spin-correlation information is lost. However, some of these effects can be undone via a method called unfolding.
Our implementation method is described in detail in the next subsection. 

\subsection{Unfolding Angular Variables} \label{sec:unfold}
As we mentioned in the last section, the detector effects distort the reconstructed observables $\tilde{A}_{LM}^i$ and $\tilde{C}_{LM,L'M'}$ from their truth-level information. To recover the underlying physics quantities that can be directly compared to generator-level predictions, these detector-level measurements should be corrected. This process, known as \textit{unfolding}, provides a systematic way to reconstruct the truth-level (parton-level) distribution from the observed detector-level one.

In this analysis, unfolding is performed for the angular observables in $h \to ZZ^* \to \ell^+\ell^- jj$ decay at a muon collider. As stated, our main goal is to obtain the distribution of the truth-level spherical harmonic coefficients $\tilde{A}_{2,0}^1$, $\tilde{C}_{2,1,2,-1}$, and $\tilde{C}_{2,2,2,-2}$, ideally free from any detector effects. This is because these coefficients are the essential inputs for estimating the quantity $\Ithree$ defined in Eq.~\ref{eq:I_3} of the $h \to ZZ^*$ system.

We correct the reconstructed distributions of the angular coefficients $\tilde{A}^{1}_{2,0}$, $\tilde{C}_{2,1,2,-1}$, and $\tilde{C}_{2,2,2,-2}$ for detector effects, using an iterative Bayesian approach based on the method of D'Agostini~\cite{DAgostini:1994fjx} and implemented via the \texttt{RooUnfold} framework\,\cite{Adye:2011gm}.
During the unfolding stage, we construct detector-to-truth response matrices and perform iterative Bayesian unfolding for the coefficients $\tilde{A}^{1}_{2,0}$, $\tilde{C}_{2,1,2,-1}$, and $\tilde{C}_{2,2,2,-2}$. Multiple \texttt{Delphes} simulated reconstructed events and their corresponding parton-level information are processed. Event selection cuts are applied to identify valid signal candidates, followed by the reconstruction of the Higgs boson kinematics and the computation of the lepton angles $(\theta, \phi)$ and jet angles $(\theta, \phi)$ in the corresponding $Z$ rest frames. From these angles, the coefficients are evaluated, and detector- and truth-level histograms are filled to build \texttt{RooUnfold} response objects for training. The trained responses are then applied to independent test datasets using the \texttt{RooUnfoldBayes} algorithm with four iterations to recover the underlying truth distributions corrected for detector effects.

\begin{figure}[htbp]
\centering
\includegraphics[width=0.32\textwidth]{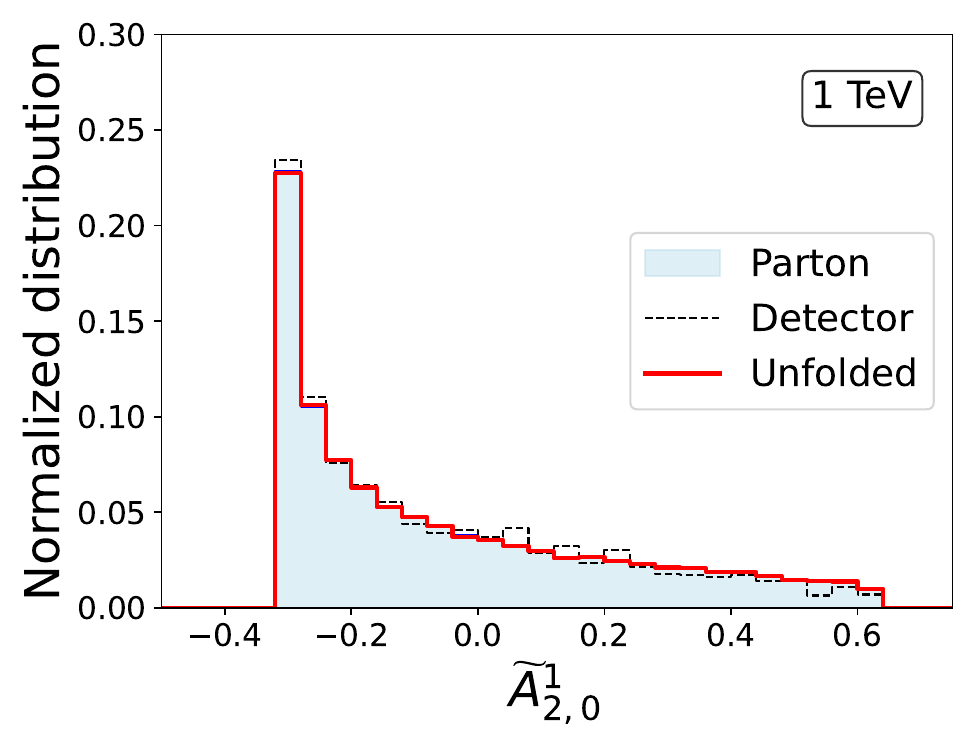}
\hfill
\includegraphics[width=0.32\textwidth]{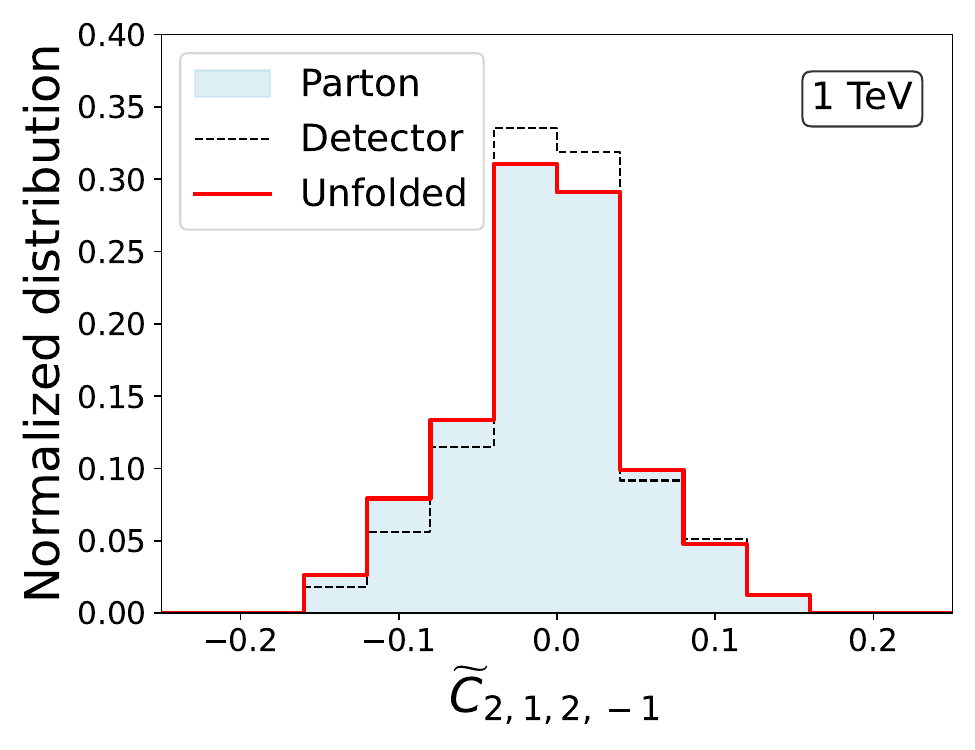}
\hfill
\includegraphics[width=0.32\textwidth]{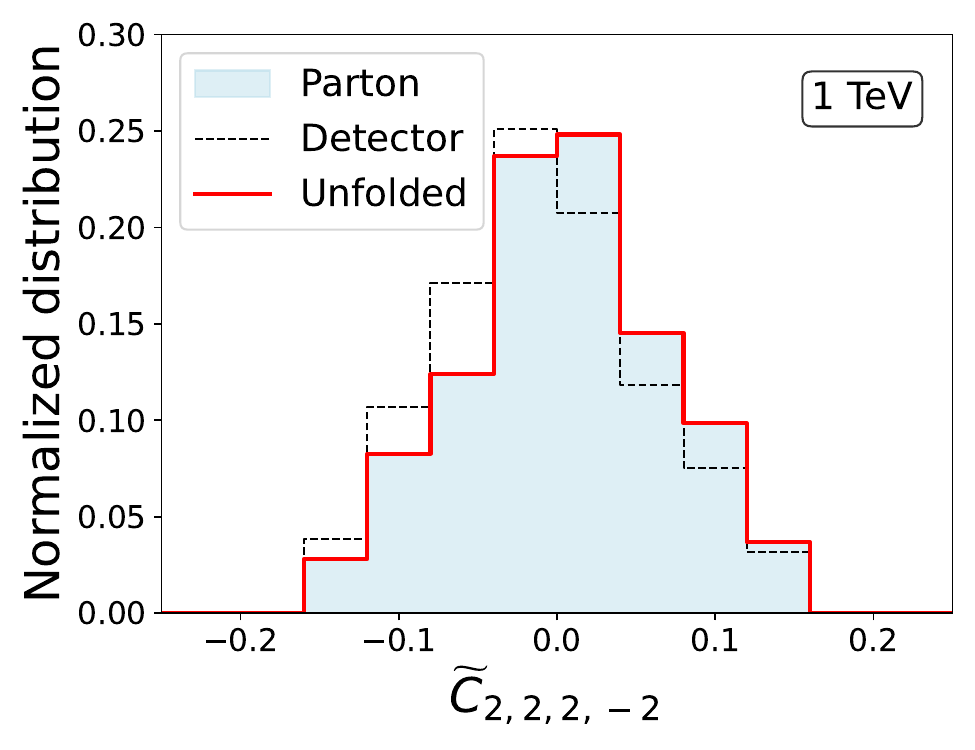}
\caption{Comparison of parton-level (filled), detector-level (dashed), and unfolded (solid red) normalized distributions for spherical harmonic coefficients at $\sqrt{s} = 1~\text{TeV}$.}
\label{fig:Unfolded_Coeffs_1TeV}
\end{figure}

\begin{figure}[htbp]
\centering
\includegraphics[width=0.32\textwidth]{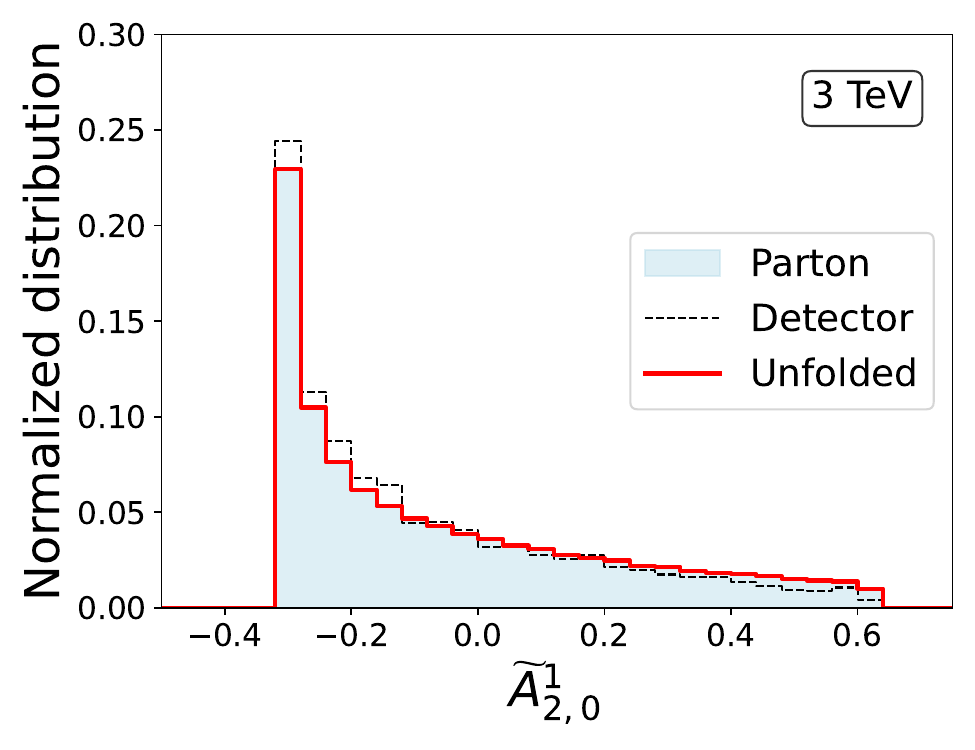}
\hfill
\includegraphics[width=0.32\textwidth]{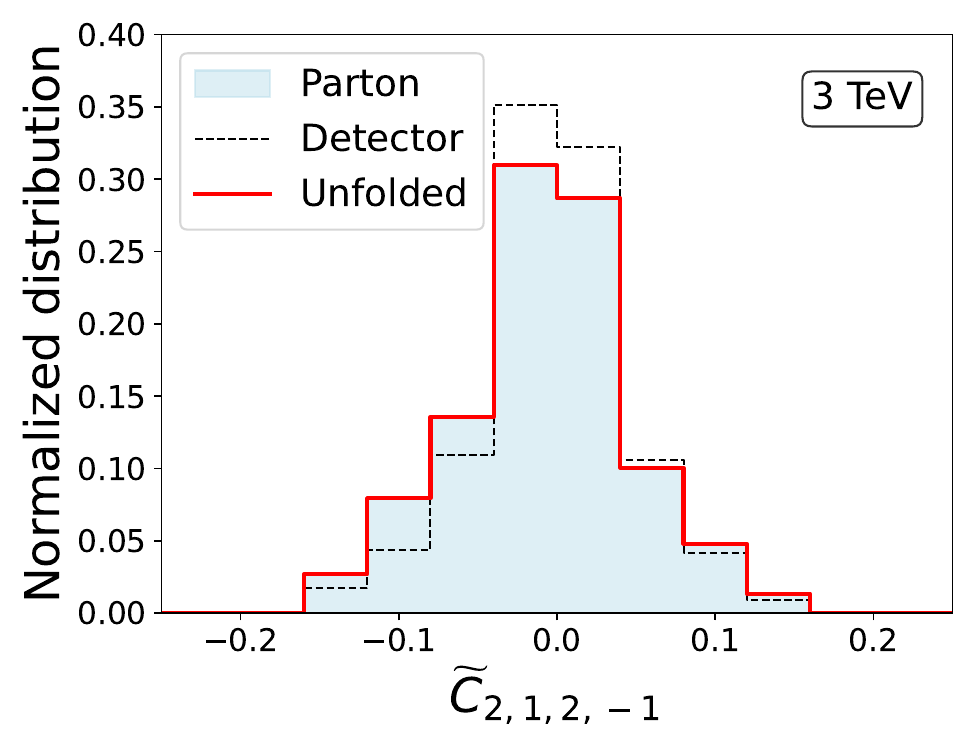}
\hfill
\includegraphics[width=0.32\textwidth]{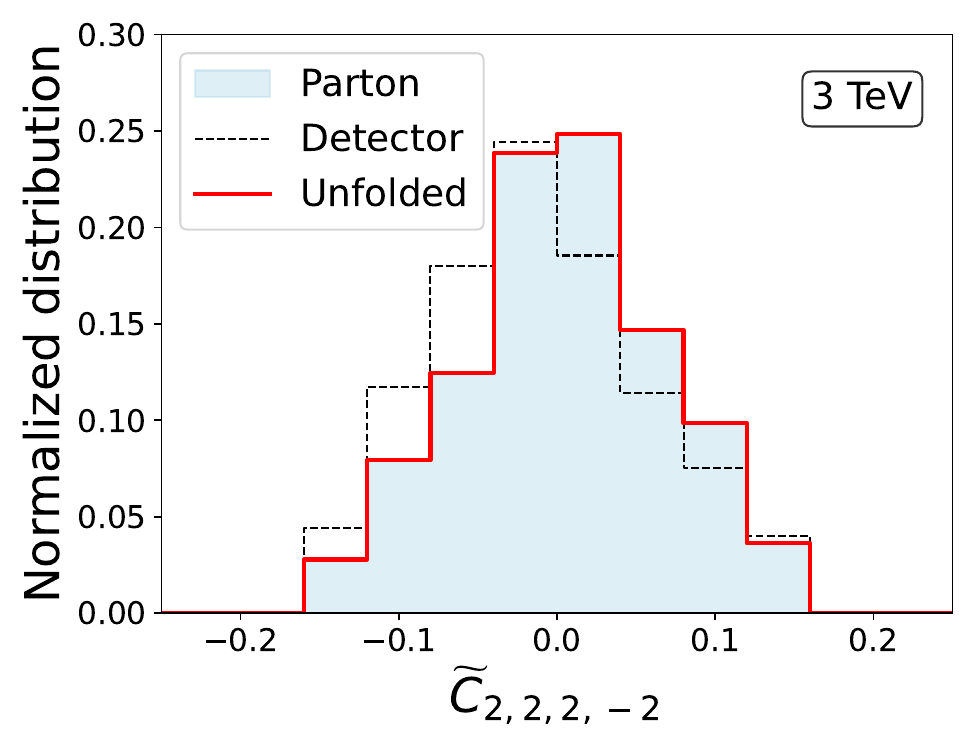}
\caption{Comparison of parton-level (filled), detector-level (dashed), and unfolded (solid red) normalized distributions for spherical harmonic coefficients at $\sqrt{s} = 3~\text{TeV}$.}
\label{fig:Unfolded_Coeffs_3TeV}
\end{figure}

\begin{figure}[htbp]
\centering
\includegraphics[width=0.32\textwidth]{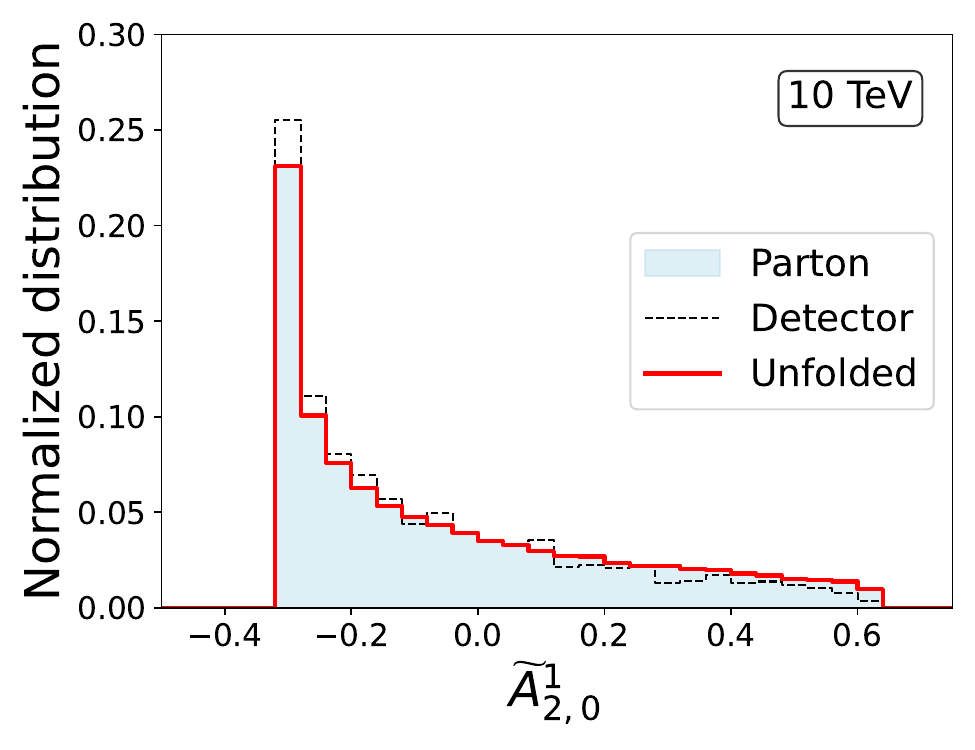}
\hfill
\includegraphics[width=0.32\textwidth]{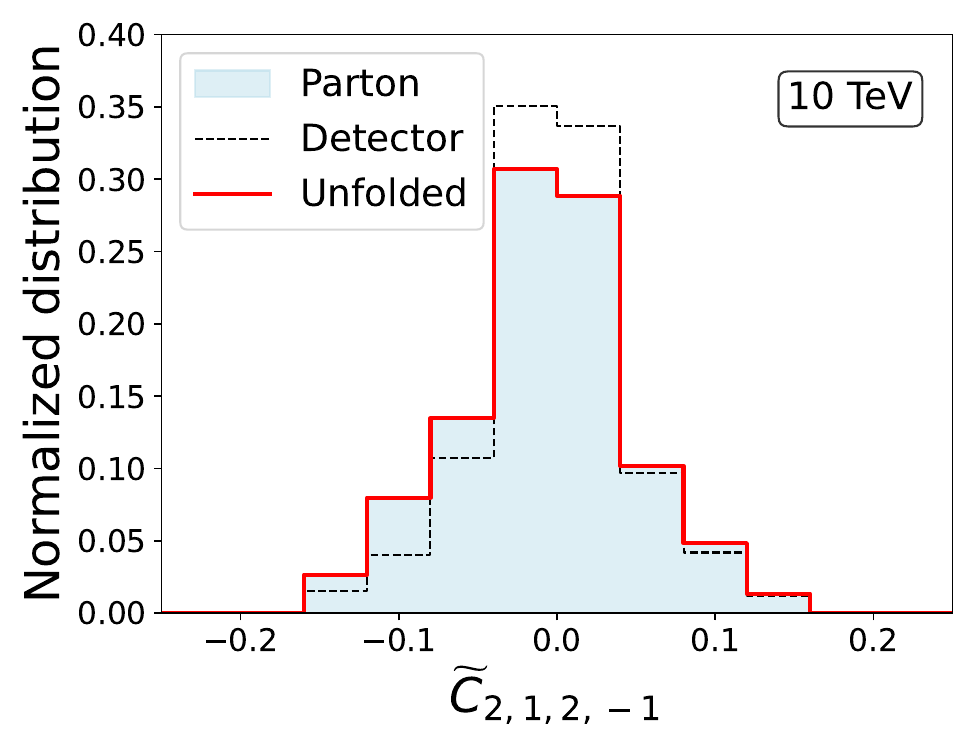}
\hfill
\includegraphics[width=0.32\textwidth]{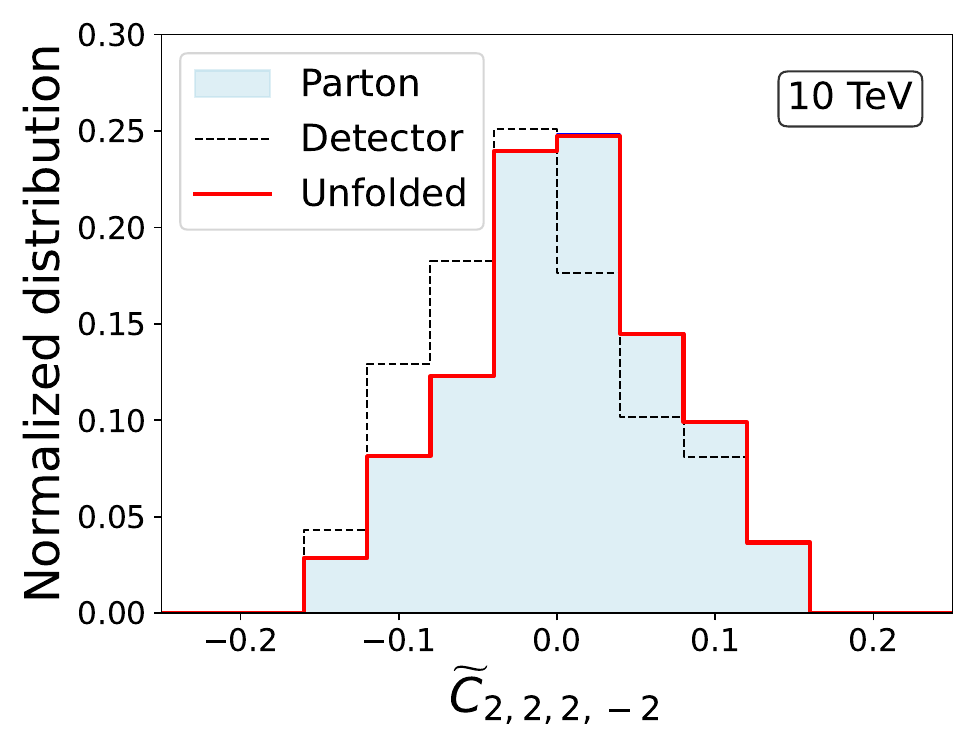}
\caption{Comparison of parton-level (filled), detector-level (dashed), and unfolded (solid red) normalized distributions for spherical harmonic coefficients at $\sqrt{s} = 10~\text{TeV}$.}
\label{fig:Unfolded_Coeffs_10TeV}
\end{figure}

The unfolding training has been performed separately for three different collider energies on the signal samples.
We show the distributions for these angular variables in Figs.~\ref{fig:Unfolded_Coeffs_1TeV}, \ref{fig:Unfolded_Coeffs_3TeV}, and \ref{fig:Unfolded_Coeffs_10TeV} for 1~TeV, 3~TeV, and 10~TeV machines, respectively, for the signal. In each figure, the left, central, and right panels, respectively, represent the distribution for $\tilde{A}^1_{2,0}$, $\tilde{C}_{2,1,2,-1}$, and $\tilde{C}_{2,2,2,-2}$. From the figures, we see that the detector-level curves (blue dashed) are distorted with respect to their parton-level (blue filled) counterparts. However, the unfolded distributions (red solid) match very well with those of the paron-level. 

An important point in this method is that we no longer have the information even-by-event. So, we have to calculate the $A$ and $C$ coefficients from the unfolded histogram. Therefore, the expressions in terms of the histogrammed bins are given by
\begin{eqnarray}
	A^i_{LM} = \frac{\sum_a u_a \tilde{A}^i_{LM,a}}{\sum_a u_a}, \qquad C_{LM,L'M'} = \frac{\sum_a v_a \tilde{C}^i_{LM,L'M',a}}{v_a}, \label{eq:ACvars}
\end{eqnarray}
where the summation runs over the bins of each histogram, $u_a$ and $v_a$ are the heights of bin $a$, and $\tilde{A}^i_{LM,a}$ and $\tilde{C}^i_{LM,L'M',a}$ are the central values of the corresponding bin. 

\section{Results}
\label{sec:results}
The results of our analysis are presented in this section, focusing on the reconstruction of the $h \to ZZ \to \ell^+\ell^- jj$ decay at a muon collider. After describing unfolding, we now move on to describe the next crucial step, which is defining a signal-enriched region that maximizes sensitivity while minimizing contamination from the SM backgrounds. 

We first apply the following `Initial' selection criteria to detector-level signal and background events:
\begin{align}
	 &{p_T}_j > 20 \; \text{GeV}, \quad {p_T}_{e,\mu} > 5.0 \; \text{GeV}, \quad |\eta_j| < 5.0, \quad |\eta_{e,\mu}| < 2.5	\nonumber\\
	 &N_j \ge 2, \qquad\qquad {\rm and}\quad (N_e = 2 \quad{\rm or}\quad N_\mu =2). \tag{Initial} \label{eq:initial}
\end{align}
Here, $N_j$, $N_e$, and $N_\mu$, respectively, represent the number of jets, electrons, muons, after imposing the restriction on the $p_T$ and $\eta$ of the respective objects. We also ensure that the lepton pairs are oppositely charged. These requirements significantly reduce the overall signal and background event samples.

\subsection{Selection of Signal Region for 1 TeV Machine}
\begin{figure}[htbp]
	\centering
\subfloat[]{\includegraphics[width=0.48\textwidth]{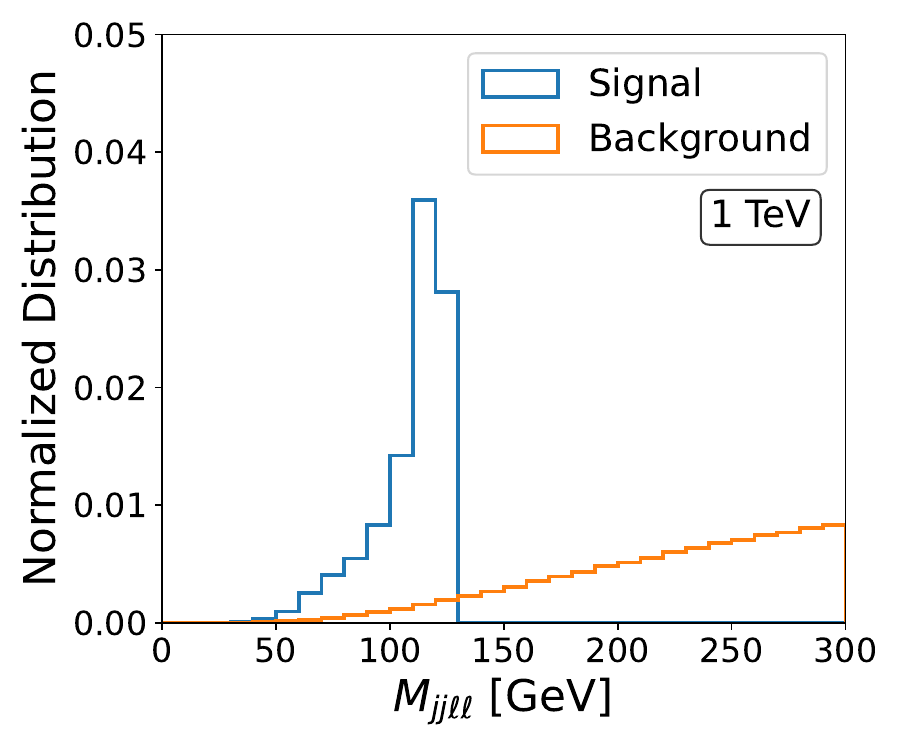}\label{fig:Mjjll_1TeV}}\hfill
\subfloat[]{\includegraphics[width=0.48\textwidth]{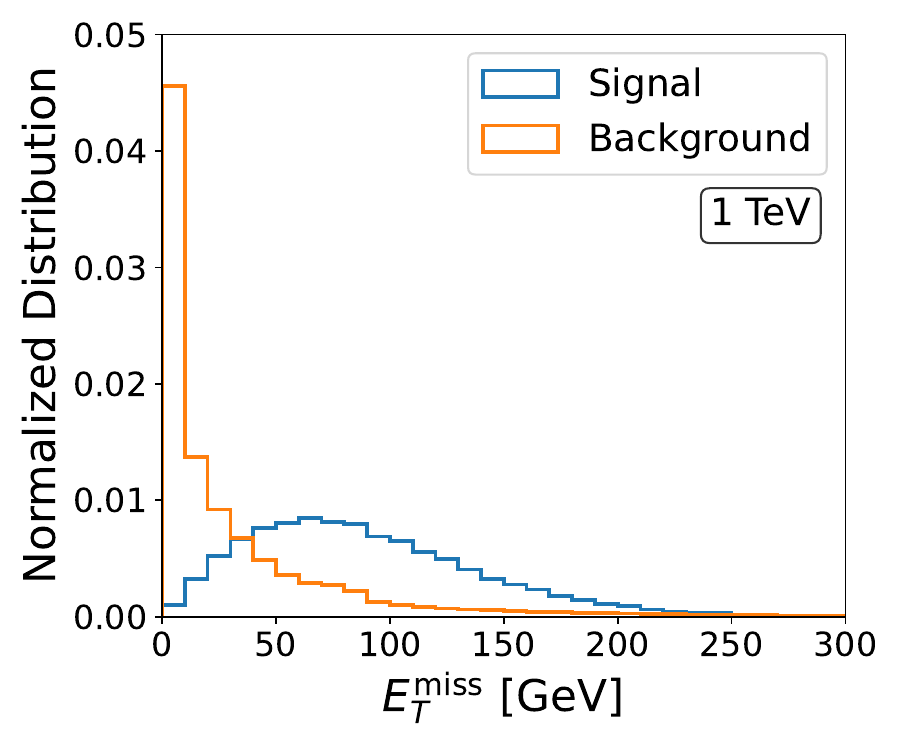}\label{fig:MET_1TeV}}
	\caption{Normalized distributions of (a) $M_{jj\ell\ell}$ and (b) the missing transverse energy $E_{T}^{\text{miss}}$ for signal and background processes at $\sqrt{s} = 1~\text{TeV}$. 
}  
	\label{fig:dist_1TeV}
\end{figure}

After imposing the `Initial' cuts on the events, we show the two-lepton-two-jet ($\ell\ell jj$) invariant mass distribution in Fig.~\ref{fig:Mjjll_1TeV}. As expected, the signal peaks at the Higgs boson mass of 125~GeV with a low mass tail due to the detector effects. On the other hand, the background has a very large invariant mass. For the missing transverse momentum shown in Fig.~\ref{fig:MET_1TeV}, the signal shows a broader $E_{T}^{\text{miss}}$ spectrum corresponding to the missing neutrino in the signal topology. The background instead peaks near low values, reflecting no direct source of missing energy in the major $\ell\ell jj$ background.
These variables, therefore, help separate the signal and reduce background events, and eventually help us select a better signal region. A cut-flow with reasonable cuts on these variables is shown in Table~\ref{Cut_1TeV}. However, even after the cuts, the number of background events remains significantly at the 1 TeV collider. Therefore, we employ a Boosted Decision Tree (BDT) based analysis for a 1~TeV machine. We now describe the BDT-based approach to optimize the signal region. The uncertainties quoted for the number of events in Table~\ref{Cut_1TeV} are calculated over 30 independent sample sets.

\begin{table}[h!]
\centering
\begin{tabular}{|l|c|c|c|c|}
\hline
                               & \multicolumn{4}{|c|}{Number of events at $\lumi =$ 10 ab$^{-1}$} \\
\cline{2-5}
Cuts                           & Signal          & $\ell\ell jj$ & $t\bar{t}$  & $WWZ$ \\
\hline
Initial                        & $500 \pm 14$ & $286\,800 \pm 700$ & $37\,920 \pm 190$ & $3\,960 \pm 40$ \\
$M_{jj\ell\ell} < 140$~GeV     & $500 \pm 14$ &     ~~~\,$47 \pm 6$   & ~~$673 \pm 27$   & ~\,$16 \pm 5$ \\
$E_{T}^{\text{miss}} > 25$~GeV & $467 \pm 14$ &     ~~~\,$24.0 \pm 3.4$ & ~~$643 \pm 25$   & ~\,$14 \pm 5$ \\
\hline
\end{tabular}
\caption{Cut flow summary for signal and background samples at detector level for $\sqrt{s}=1$~TeV at \lumi = 10~ab$^{-1}$.}
\label{Cut_1TeV}
\end{table}

\subsubsection{Boosted Decision Tree Analysis for 1 TeV Collider.}
Boosted Decision Trees are widely used multivariate techniques in high-energy physics for event classification. A BDT is an ensemble of decision trees, 
based on the input of many variables.
This framework effectively allows us to capture complex correlations between input variables, thereby helping us select the signal region optimally. We use the publicly available implementation Extreme Gradient Boosting ({\tt XGBoost})\,\cite{Chen:2016btl} for the BDT classifier. In our analysis, the BDT is trained using simulated signal events ($h \to ZZ \to \ell^+\ell^- jj$) as one class to be distinguished from the other class represented by the background events. For input features, we have used the following variables: 
\begin{itemize}
    \item Invariant mass of the final states, $M_{\ell\ell jj}$,
    \item Missing transverse energy, $E_{T}^{\text{miss}}$,
    \item $\Delta R$ between the two leading $p_T$ jets, $\Delta R_{jj}$,
    \item $\Delta R$ between the two leading $p_T$ leptons, $\Delta R_{\ell\ell}$,
    \item Transverse momenta of the two leading $p_T$ jets, $p_T^{j_1}$ and $p_T^{j_2}$,
    \item Transverse momenta of the two leading $p_T$ leptons, $p_T^{\ell_1}$ and $p_T^{\ell_2}$.
\end{itemize}
These variables consist of the global event kinematics and the internal correlations between leptons and jets, providing the BDT with sufficient information to efficiently separate signal from background events. 

\begin{figure}[htbp]
\centering
\subfloat[]{\includegraphics[width=0.44\textwidth]{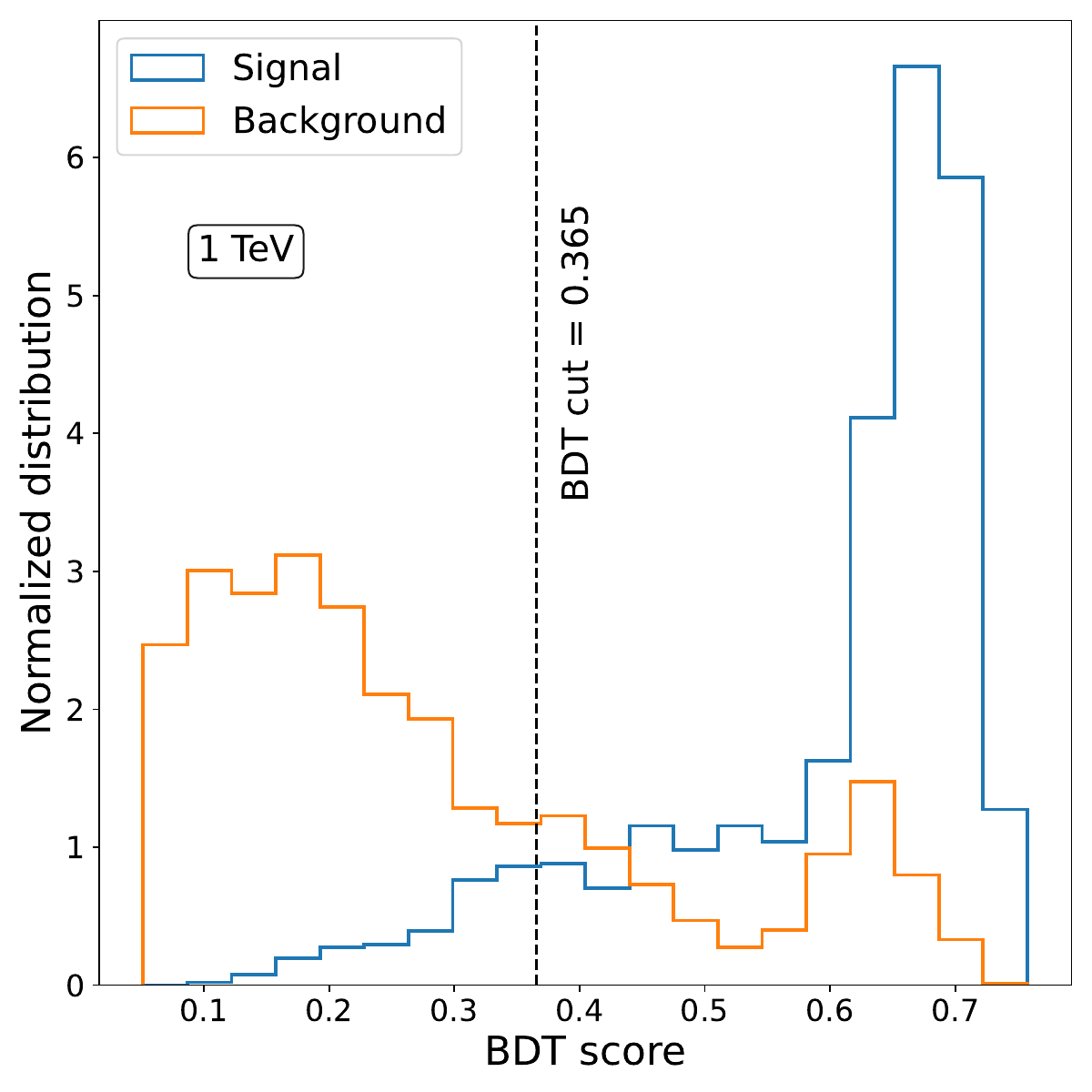}\label{fig:BDTclassifier}}	\hfill
\subfloat[]{\includegraphics[width=0.53\textwidth]{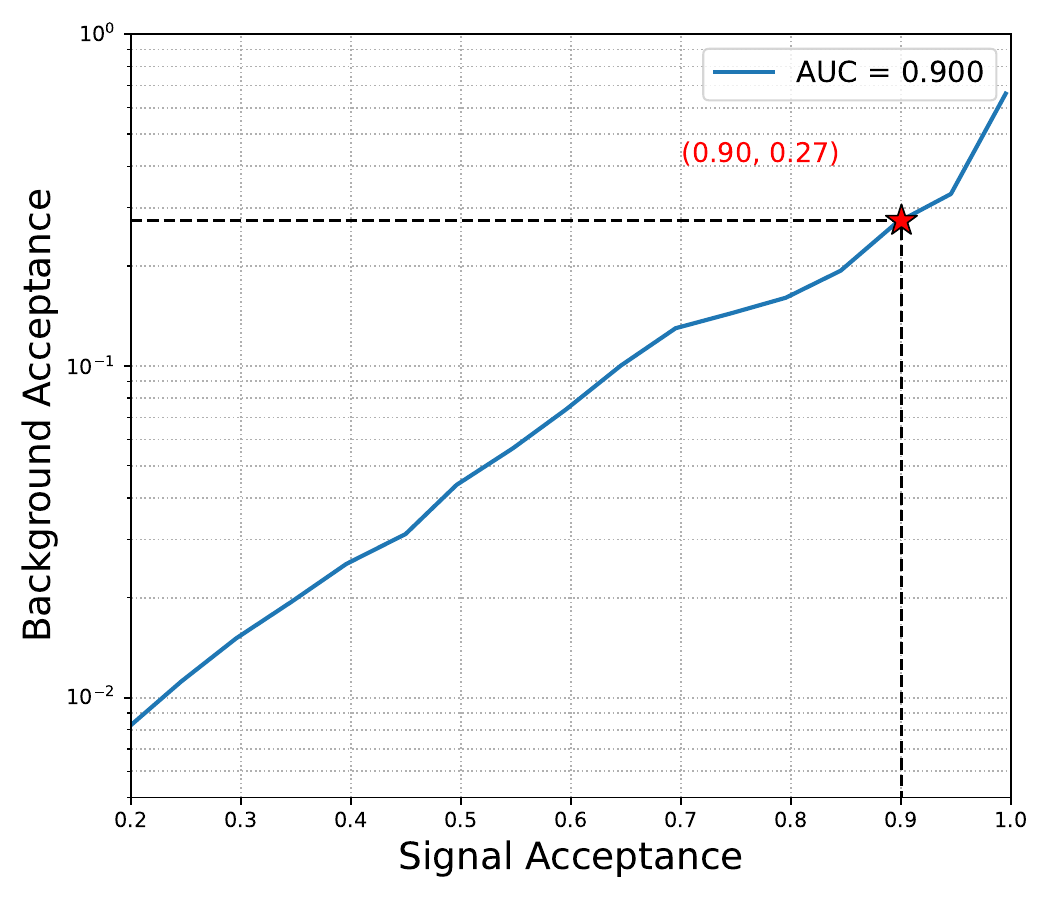}\label{fig:ROCcurve}}
\caption{(a) Normalized BDT score distributions for signal and background samples.
The vertical dashed line at $\text{BDT} = 0.365$ indicates the optimized selection threshold used for the analysis. (b) Receiver Operating Characteristic (ROC) curves for the BDT classifier. The point marked by $*$ represents the selection threshold.
The area under the curve (AUC) indicates strong discriminating power.}	
\label{fig:BDTscores}
\end{figure}

An advantage of the BDT framework is that it combines information from more than one input variable to provide a BDT classifier score that helps distinguish signal from background. The distributions of the BDT score, with a specific choice of {\tt XGBoost} model parameters, are shown in Fig.~\ref{fig:BDTclassifier}. As we can see, the distribution is quite separate for the signal from the background. One can then use an optimal threshold on this BDT score to isolate the signal region. 
For a given threshold selected, there are signal and background acceptances. A measure of goodness of the BDT classifier is given by the receiver operating characteristic (ROC) curve, which is the curve of signal acceptance versus background acceptance by varying the threshold. We show the ROC curve for our BDT classifier in Fig.~\ref{fig:ROCcurve}, and the area under the curve is 90\%, showcasing a good performance of the network.

We train the BDT network with the events selected after imposing $M_{jj\ell\ell}$ and $E_{T}^{\text{miss}}$ cut as given in Table~\ref{Cut_1TeV}. We perform a scan over the XGBoost model parameters to choose an optimal set of hyperparameters. We have also performed overtraining checks by comparing the BDT score distributions and ROC curves for the training and independent validation sets in selecting the optimal hyperparameter choices. 

We then choose the BDT score threshold at 0.365, corresponding to signal acceptance of 90\% and background acceptance of 27\% for the signal region. The events above this threshold are selected to belong to the signal-enriched region. Subsequently, the unfolding of the $\tilde{A}$ and $\tilde{C}$ observables is performed on these selected events. The optimally chosen signal region ensures reduced background contamination and, therefore, better reconstruction of the angular observables after unfolding. We then calculate $\Ithree$ using these unfolded distributions by the method described in Eq.~(\ref{eq:ACvars}) in Section~\ref{sec:unfold}. 

\subsection{Signal Region Selection for 3 and 10 TeV Machine}
\begin{figure}[htbp]
\hfill
\subfloat[]{\includegraphics[width=0.48\textwidth]{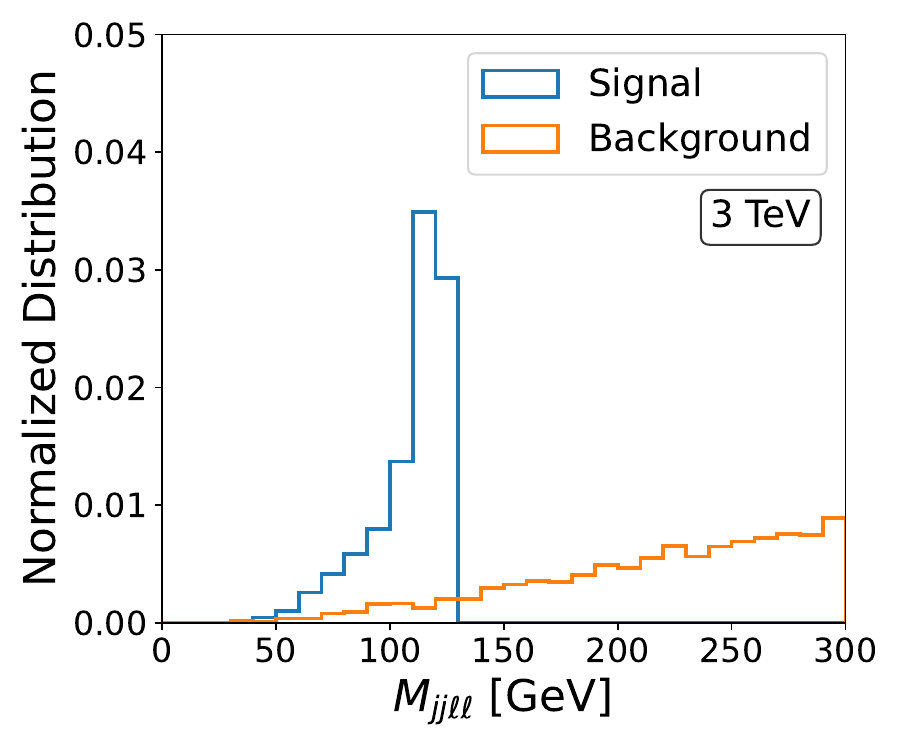}}
\hfill
\subfloat[]{\includegraphics[width=0.48\textwidth]{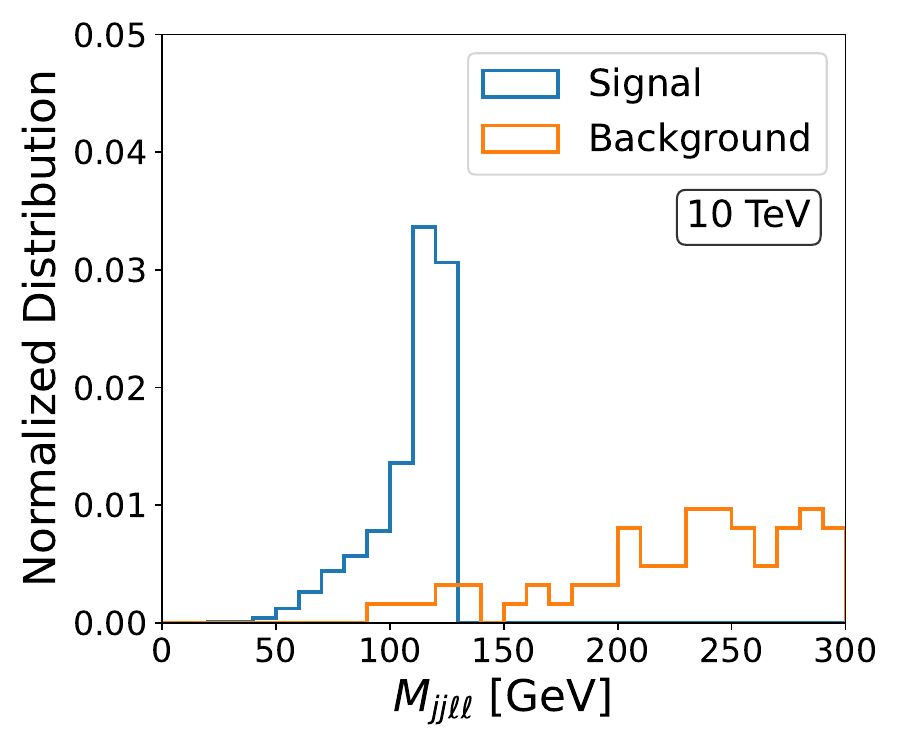}}
\caption{Normalized distributions of the invariant mass $M_{jj\ell\ell}$ for signal and background events at (a) 3 TeV and (b) 10 TeV muon collider. The blue histograms represent the signal, while the orange ones correspond to the combination of all background events.}
\label{fig:Mjjll_all}
\end{figure}

The backgrounds are fewer in the 3 and 10 TeV machines (see Table~\ref{events}), unlike the 1 TeV case. The signal can be isolated from the backgrounds relatively easily by a simple cut on $M_{jj\ell\ell}$. To demonstrate that, we first show the distribution of $M_{jj\ell\ell}$ variable in Fig.~\ref{fig:Mjjll_all} after imposing the `Initial' cuts in Eq.~\ref{eq:initial}. As can be seen from the distributions, the dilepton-dijet invariant mass peaks at $M_h = 125$ GeV, whereas the backgrounds are distributed evenly over the range. We, therefore, set an upper cut on the variable $M_{jj\ell\ell} < 140$ GeV, which reduces the background significantly. This cut flow, along with expected numbers of events after applying each cut, is given in Tables~\ref{Cut_3TeV} and \ref{Cut_10TeV} for the 3 and 10 TeV machines, respectively.

For a 3 TeV machine with 10~ab$^{-1}$ luminosity, the expected number of background events reduces to less than 2\% of the expected signal events. For a 10 TeV machine with the same luminosity, the expected number of background events is negligible, indicating an almost pure signal region. 

\begin{table}[h!]
\centering
\begin{tabular}{|l|c|c|c|c|}
\hline
Cuts & Signal & $\ell\ell jj$ & $t\bar{t}$ & $WWZ$ \\
\hline
Initial                    & $852 \pm 30$ & $81\,540 \pm 270$ & $2\,650 \pm 40$ & $780 \pm 40$ \\
$M_{jj\ell\ell} < 140$~GeV & $852 \pm 30$ & ~~~\,$1.3 \pm 0.6$        & ~~$13.0 \pm 2.9$    & $< 1$ \\
\hline
\end{tabular}
\caption{Cut flow summary for signal and background samples at detector level for $\sqrt{s}=3$~TeV.  }
	\label{Cut_3TeV}
\end{table}

\begin{table}[h!]
	\centering
	\begin{tabular}{|l|c|c|c|c|}
		\hline
		Cuts & Signal & $\ell\ell jj$ & $t\bar{t}$ & $WWZ$ \\
		\hline
		Initial & $1\,010 \pm 40$ & $16\,700 \pm 120$ & $110 \pm 10$ & $90 \pm 6$ \\
		$M_{jj\ell\ell} < 140$~GeV & $1\,010 \pm 40$ & $< 1$ & $< 1$ & $< 1$ \\
		\hline
	\end{tabular}
\caption{Cut flow summary for signal and background samples at detector level for $\sqrt{s}=10$~TeV. }
	\label{Cut_10TeV}
\end{table}
\subsection{Estimation of $\Ithree$} \label{sec:I3estm}
After selecting the signal-enriched phase-space regions, we are ready to compute the value of the CGLMP measure $\Ithree$ using Eq.~\ref{eq:I_3}. To remind the reader, the signal regions are defined by the BDT cut along with `\ref{eq:initial}' requirements, and by the $M_{jj\ell\ell}$ and $E_{T}^{\text{miss}}$ requirements listed in Table~\ref{Cut_1TeV} for the 1 TeV machine. For the 3 and 10 TeV machines, the signal regions are defined by the same `\ref{eq:initial}' cuts together with the $M_{jj\ell\ell}$ selection requirements summarized in Tables~\ref{Cut_3TeV} and \ref{Cut_10TeV}, respectively. Estimates of $\Ithree$ are performed on 30 independent Monte Carlo samples for the signal and the backgrounds, where each sample contains the number of events expected at 10~ab$^{-1}$, as listed in Table~\ref{events}. The unfolding procedure, described in Section~\ref{sec:unfold} and trained separately on the 1 TeV, 3 TeV, and 10 TeV signal samples, is applied when calculating $\Ithree$. We then calculate the mean and statistical uncertainties of $\Ithree$ from these samples.

\begin{table}[htbp]
\centering
\begin{tabular}{|c|c|c|c|}
\hline
\multirow{2}{*}{$M_{Z^*}$ [GeV]} & \multicolumn{3}{c|}{$\Ithree$} \\
\cline{2-4}
& $\sqrt{s} = 1~\text{TeV}$ & $\sqrt{s} = 3~\text{TeV}$ & $\sqrt{s} = 10~\text{TeV}$ \\
\hline \hline
[0, 10)   & 2.629 $\pm$ 0.080  & 2.614 $\pm$ 0.033 & 2.580 $\pm$ 0.033 \\
\hline
~\,[10, 20)  & 2.637 $\pm$ 0.031  & 2.626 $\pm$ 0.014 & 2.569 $\pm$ 0.019 \\
\hline
~\,[20, 30)  & 2.613 $\pm$ 0.028  & 2.627 $\pm$ 0.012 & 2.569 $\pm$ 0.017 \\
\hline
~\,[30, 40)  & 2.600 $\pm$ 0.052  & 2.620 $\pm$ 0.017 & 2.579 $\pm$ 0.029 \\
\hline
\hline
\end{tabular}
\caption{Expected values of $\Ithree$ in different $M_{Z^*}$ bins for $\sqrt{s}=1$, 3, and 10~TeV, assuming 10 ab$^{-1}$ integrated luminosity for each. The values include only the contribution from the signal within the signal region.}
\label{tab:I3-mz2}
\end{table}

\begin{figure}[htbp]
\centering
\includegraphics[width=0.65\textwidth]{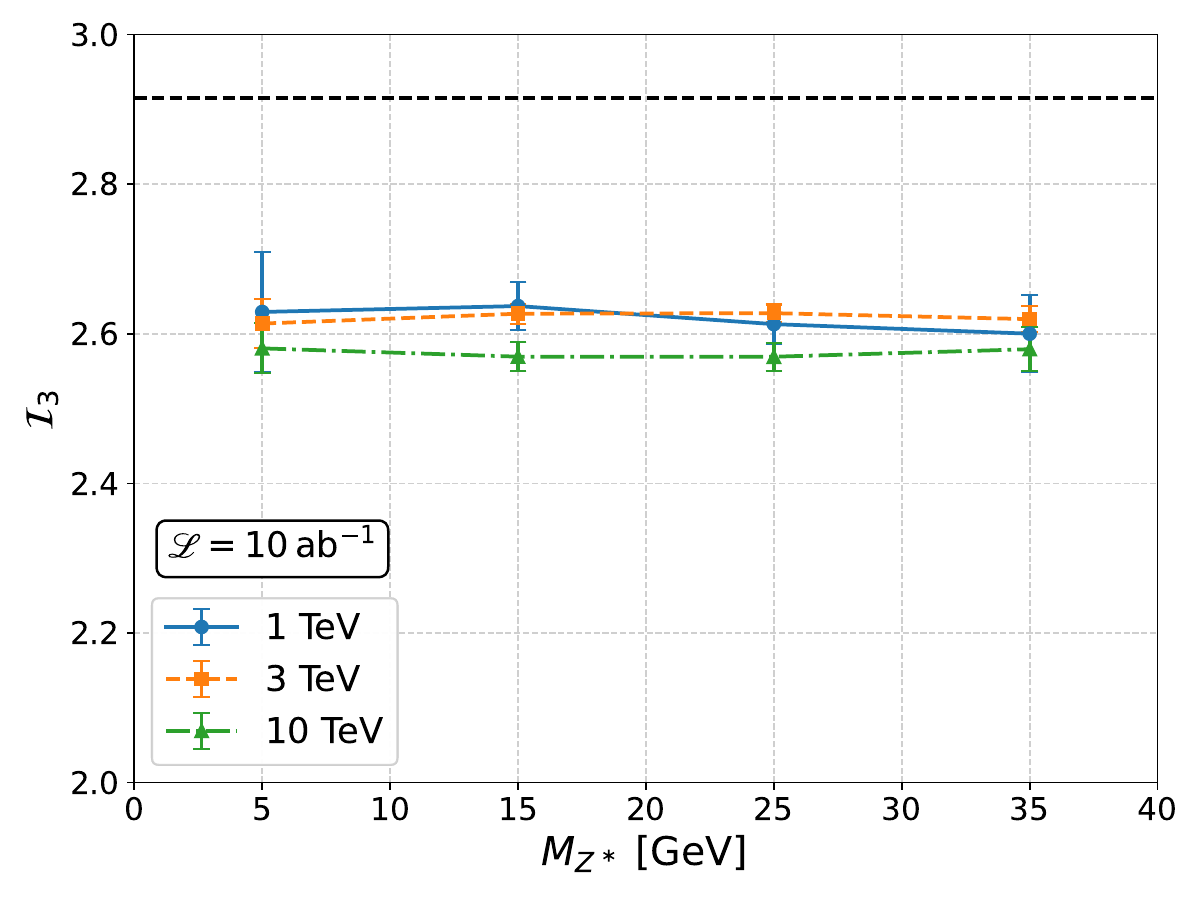}
\caption{Variation of the CGLMP observable $\Ithree$ as a function of $M_{Z^*}$ for different center-of-mass energies. The plot shows results for $\sqrt{s} = 1~\text{TeV}$ (blue solid), $3~\text{TeV}$ (orange dashed), and $10~\text{TeV}$ (green dash-dotted). The error bars represent the statistical uncertainties for each point. The horizontal dashed line represents the maximum achievable value for $\Ithree = 2.9149$ for a two-qutrit system.}
\label{fig:I3_vs_MZp}
\end{figure}

The degree of off-shellness of the $Z^*$ contributes to different levels of involvement of the longitudinal mode, resulting in different levels of entangled states in the system\,\cite{Aguilar-Saavedra:2022wam}. We first present in Table~\ref{tab:I3-mz2}, the mean values of $\Ithree$ for different choices of $M_{Z^*}$ for 1, 3, and 10~TeV muon colliders, assuming an integrated luminosity of 10 ab$^{-1}$. These values are plotted in Fig.~\ref{fig:I3_vs_MZp}, with error bars representing the corresponding statistical uncertainties. 
The $x$-values in Fig.~\ref{fig:I3_vs_MZp} denote the midpoint of each $M_{Z^*}$ bin. The statistical uncertainties on the expected $\Ithree$ are very small. This is made possible by the high event statistics in the hadronic final state.
As we can see, the mean value of $\Ithree$ is around 2.6, close to the maximal achievable value of 2.9149 in a qutrit-qutrit system\,\cite{Acin:2002zz}. This indicates that a high level of quantum entanglement can be accessed. For these estimations, only signal events within the signal region are included.

\begin{table}[h]
\centering
\begin{tabular}{c p{0.1\textwidth} c}
	\hline\hline
	$\mathbf{E_{\rm CM}}$ &  &  $\Ithree$ \\ \hline
	1 TeV &   & $2.625\pm 0.020$ \\ 
	3 TeV &   & $2.623\pm 0.007$ \\ 
	10 TeV &   & $2.582\pm 0.017$ \\
	\hline\hline
\end{tabular}
\caption{Expected values of $\Ithree$ over full $M_{Z^*}$ range for $\sqrt{s}=1$, 3, and 10~TeV, assuming 10 ab$^{-1}$ integrated luminosity for each machine. The values include the contributions from both signal and background within the signal region.}
\label{tab:I3vals}
\end{table}
Next, we present the values of $\Ithree$ in the full $M_{Z^*}$ range for the three different machine energies, each with 10~ab$^{-1}$ integrated luminosity, in Table~\ref{tab:I3vals}. The values are computed using both signal and background events in the signal region since backgrounds cannot be completely eliminated. The unfolding procedure, trained only on the signal samples, is applied to both the signal and background contributions. The expected values of $\Ithree$ are found to be $2.625\pm 0.020$, $2.623\pm 0.007$, and $2.582\pm 0.017$ for 1 TeV, 3 TeV, and 10 TeV muon colliders, respectively. Once again, the expected $\Ithree$ values reach as high as 2.625 with very small uncertainties, thanks to the large statistics in the hadronic channel. A high level of entanglement becomes accessible because the unfolding procedure effectively corrects for the detector and hadronization effects in the hadronic states.

These results demonstrate that quantum entanglement in the $h\to ZZ^*$ system can be probed with high precision at future muon colliders, even in the presence of backgrounds, hadronization, and detector effects. The consistently large values of $\Ithree$ indicate that the hadronic decay mode, supported by unfolding, provides a robust technique to observe strong entanglement signatures at high energies.

\section{Summary and Outlook}\label{sec:conclusion}
Test of quantum entanglement, an emergent feature of quantum theory, has received considerable attention in recent years. In this work, we have performed such a study of quantum entanglement through the violation of Bell-type inequalities in the $h \to Z Z^*$ system at a future muon collider. We choose the muon collider because of its relatively clean environment and the possibility of achieving higher centre-of-mass energy and integrated luminosity. Our analysis focuses on the opposite-sign dilepton plus dijet final state, where one $Z$ boson is produced on-shell, and the other remains off-shell.

We adopt the generalized CGLMP construction as an entanglement measure with the Bell operator that maximizes the extraction of the quantum correlations. The CGLMP measure for this system can be expressed in terms of the spin correlation angular observables of the decay products of the $Z^{(*)}$ bosons, which in turn are reconstructible from the kinematics of the dilepton-dijet final state. We have performed the study, including detector effects using fast detection simulation from {\tt Delphes}. We then correct the smearing from the hadronization and detector effects on the hadronic final states using an unfolding method based on {\tt RooUnfold}.

The analysis has been performed at three different muon collider energies: 1, 3, and 10~TeV. At a 1 TeV collider, due to a low signal-to-background ratio, a BDT-based multivariate analysis is used to isolate a signal-enriched region. At 3 and 10 TeV, the signal can be efficiently selected by a simple upper cut on the dilepton-dijet invariant mass. After selecting the signal, the trained unfolding model is applied to correct the angular observables, which are then used to compute the CGLMP measure $\Ithree$. Our predictions are $\Ithree =  2.625\pm0.012$, $\Ithree = 2.623\pm0.004$, and $\Ithree = 2.582\pm0.010$ for 1, 3, and 10 TeV, respectively, assuming 10 ab$^{-1}$ luminosity. These values show that a high level of quantum entanglement, close to the maximal theoretical value of $2.9149$, can be measured with very small uncertainties, primarily due to the large event yield in the hadronic states.

Although the present study is performed in the relatively clean environment of a lepton collider, our unfolding-based strategy provides a path toward carrying out similar quantum correlation studies even at hadron colliders, where detector and QCD effects are far more severe. An immediate extension of this work would be a dedicated analysis of the fully hadronic $Z Z^* \to 4j$ channel, which offers much higher statistics but also brings stronger hadronization and jet-combinatorial challenges. It would also be interesting to explore other measures of entanglement and alternative Bell-type operators that may provide enhanced sensitivity to the underlying quantum correlations. On the technical side, the unfolding procedure can be further improved by adopting more advanced regularized techniques or machine learning-based unfolding methods, such as neural-network or normalizing-flow approaches, which could yield more accurate reconstruction of the multi-dimensional angular observables. Overall, our study shows that quantum entanglement in Higgs decays can be effectively extracted even in challenging hadronic final states, and it opens up several promising directions for carrying out more refined and realistic investigations of quantum correlations at future high-energy colliders.

\section*{Data Availability Statement}
This work uses simulated data generated with publicly available software packages. The simulations and analysis procedures are fully described in this article.

\acknowledgments
The authors acknowledge the High-Performance Computing facility Nandadevi at The Institute of Mathematical Sciences for supporting the computational needs of this work. A.D. acknowledges the support from Carl Trygger Foundation (grant CTS23-2930). T.S.~acknowledges the fruitful discussions with Prof. Pankaj Agrawal and Dr. Biswajit Das. S.D. thanks the Theoretical Physics group at IMSc, Chennai, for their hospitality, where most of this work was completed.


\providecommand{\href}[2]{#2}\begingroup\raggedright\endgroup

\end{document}